\documentclass{aa}  

\usepackage{graphicx}
\usepackage{subfig}
\usepackage{placeins}
\usepackage{txfonts}
\usepackage{comment}
\usepackage{amstext}
\usepackage[normalem]{ulem}

\begin{document}

 \title{Stellar population gradients at cosmic noon as a constraint to the evolution of passive galaxies}
   \author{F. R. Ditrani\inst{1}\inst{2}, S. Andreon\inst{2}, M. Longhetti\inst{2}, A. Newman\inst{3}}
          
   \institute{Università degli studi di Milano-Bicocca, Piazza della scienza, I-20125 Milano, Italy\\
   \email{f.ditrani1@campus.unimib.it}
   \and
   INAF-Osservatorio Astronomico di Brera, via Brera 28, I-20121 Milano, Italy\\
        \and Carnegie Institution for Science, 813 Santa Barbara St., Pasadena, California 91101, USA
             }

   \date{Received; accepted }

  \abstract
    {The radial variations of the stellar populations properties within passive galaxies at high redshift contain information about their assembly mechanisms, based on which galaxy formation and evolution scenarios may be distinguished.}
   {The aim of this work is to give constraints on massive galaxy formation scenarios through one of the first analyses of age and metallicity gradients of the stellar populations in a sample of passive galaxies at $z > 1.6$ based on spectroscopic data from the Hubble Space Telescope.}
  {We combined G$141$ deep slitless spectroscopic data and F$160$W photometric data of the spectroscopically passive galaxies at $1.6< z < 2.4$ with $H_{160} < 22.0$ in the field of view of the cluster JKCS\,$041$. We extracted spectra from different zones of the galaxies, and we analysed them by fitting them with a library of synthetic templates of stellar population models to obtain estimates of the age and metallicity gradients.}
  {We obtained reliable measurements of age and metallicity parameters in different spatial zones of $\text{four}$ galaxies. We performed spatially resolved measurements in individual high-redshift galaxies without the need of peculiar situations (i.e. gravitational lensing) for the first time. All four galaxies exhibit negative metallicity gradients.
  Their amplitude, similar to that measured in galaxies in the local Universe, suggests 
  that the stellar populations of  passive galaxies from $z \sim 2$ to $z = 0$ are not redistributed.}
  {Although the sample we analysed is small, the results we obtained suggest that the main mechanism that determines the spatial distribution of the stellar population properties within passive galaxies is constrained in the first $3$ Gyr of the Universe. This is consistent with the revised monolithic scenario.}
   {}

   \keywords{galaxies: clusters: individual (JKCS\,$041$) -- galaxies: evolution -- galaxies: high-redshift -- galaxies: formation -- galaxies: stellar content
               }
\titlerunning{Stellar population gradients at cosmic noon}
\authorrunning{Ditrani et al.}
   \maketitle
%

\section{Introduction}
\label{sec:intro}
    Understanding the formation and evolution of early-type galaxies (ETGs) is important because they contain almost $80\%$ of the baryonic mass of the local Universe \citep[][and references therein]{renzini2006stellar}. ETGs dominate the highest-density regions, such as clusters of galaxies \citep{dressler1980galaxy,dressler1984evolution,balogh1999differential} up to $z = 1.8$ \citep{raichoor2012star,andreon2014jkcs,strazzullo2019galaxy,willis2020spectroscopic}. In local clusters, ETGs are mainly elliptical and spheroidal galaxies, and they are populated by almost coeval old stars that apparently descended from a single star formation episode that occurred at $z>2-3$ \citep[e.g.][]{thomas2010environment}. 
    In the past decade, deep-field and cluster photometric surveys \citep[e.g.][]{kurk2009gmass,van2010growth,papovich2010spitzer,croom2012sydney,cassata2013constraining,straatman2014substantial} and spectroscopic studies \citep[e.g.][]{kriek2009ultra,gobat2012early,van2013stellar,newman2014spectroscopic,andreon2014jkcs,morishita2019massive,willis2020spectroscopic} observed massive and passive galaxies ($M > 10^{11}$ M$_\odot$) that were already in place at $z \sim 2-3$, and their stars were old compared with the age of the Universe at these redshifts.
    
    Although numerous samples of passive galaxies have been analysed in the past decade, the mechanisms that describe their formation and evolution are still unclear. In particular, the relative importance of the environment and the mass in the regulation of the formation and the evolution of the ETGs is still debated \citep[e.g.][]{treu2003wide,thomas2010environment,raichoor2012star,2021arXiv211002860F}. In rich clusters, the environment can play an important role in the evolution of ETGs. Various processes can regulate the ETG life in dense environments, such as interactions of the galaxy with the intracluster medium (ICM) \citep[e.g. ram-pressure stripping;][]{gunn1972infall,fujita1999effects}, galaxy-cluster gravitational interactions \citep[e.g. tidal interactions;][]{byrd1990tidal,henriksen1996tidal}, and interactions between galaxies \citep[e.g. mergers;][]{icke1985distant,bekki1998unequal}.
    
    Several theoretical scenarios have been proposed to describe ETG formation and assembly.
    In the hierarchical galaxy formation model \citep{cole1994recipe,baugh1996evolution,kauffmann1996age,kauffmann1998chemical},  ETGs are formed through subsequent wet mergers of pre-existing small galaxies moving in the same potential well  \citep{toomre1972galactic,de2006formation}. In clusters, the mass assembly takes place earlier than in the field, with a higher rate of mergers \citep[e.g.][]{maulbetsch2007dependence,de2007hierarchical}. ETGs grow faster in clusters than in the field \citep{andreon2018cosmic}.
     
    Alternatively, ETGs might assemble their mass at $z >2-3$ through the merger of small substructures that move in a common potential well \citep{dekel2009cold} in the so-called revised monolithic scenario.
    An almost purely passive evolution is expected to follow the strong initial activity, and only weak episodes of star formation at $z < 1$ are foreseen, for example caused by the capture of small satellites \citep{katz1991dissipational,kawata2001role,kobayashi2004grape,merlin2006formation}.
    
    In recent years, the inside-out growth model has become widely accepted. Supported by several simulations \citep[e.g.][]{khochfar2006simple,hopkins2009compact,wuyts2010sizes,naab2009minor,bezanson2009relation} and by observations of ETGs at $1.0 < z < 2.5$ with an effective radius that is $3-5$ times smaller than the mean radius of the local ETGs with the same mass \citep[e.g.][]{newman2012can,newman2014spectroscopic,andreon2016size,strazzullo2019galaxy}, the model assumes that passive galaxies assemble as compact spheroids through wet merger events at $z > 4-5$. At lower redshift, the spheroids undergo various dry mergers with low-mass objects, which increases their effective radius.
    
    These scenarios can be distinguished through a spatially resolved analysis of the stellar populations properties within passive galaxies. While in the hierarchical scenario, flat profiles of stellar population properties are expected \citep{bekki1999stellar,naab2009minor}, in the monolithic scenario, stellar populations are expected to be more metal rich in the core zone of the galaxies than in their outskirts. Moreover, stochastic gradients are expected at different redshift in the inside-out growth scenario, where the final gradient depends on the nature and time of the satellite accretion events. 
    Therefore, different distributions of stellar population properties within passive galaxies indicate different formation and/or evolution processes. Moreover, measurements of age and metallicity gradients in high-redshift galaxies, that is, in an epoch closer to their formation, can provide more direct constraints on the assembly mechanisms predicted by the models described above.

    Recent studies that have analysed several samples of massive ETGs at low redshift \citep[e.g.][]{kelson2006line,spolaor2009mass,spolaor2010early,koleva2011age,scott2009sauron,greene2015massive,ferreras2019sami,zibetti2020insights}, in particular in the Coma and Fornax clusters \citep[][and references therein]{mehlert2003spatially,2006A&A...457..823S,bedregal2011stellar} found strong negative metallicity gradients. The situation is less clear for the age gradients. Some studies show mild or null age gradients, while other studies reported strong positive age gradients \citep[e.g.][]{kuntschner2010sauron,zibetti2020insights}. 
    It is important to note that age gradients of only $1-2$ Gyr in local ETGs are difficult to detect in stellar popoulations that are more than $10$ Gyr old on average, while their impact on the observed spectrophotometric properties is thought to be much more evident at higher redshift. 
    
    The study of age and metallicity gradients of the stellar populations in passive galaxies at high redshift would overcome the difficulty of detecting small age differences in old stellar populations and would offer information on the star formation history (SFH) and the mass assembly of ETGs closer to their formation epoch.
    Moreover, the age-metallicity degeneracy is predominant at low redshift \citep{worthey1994old}. 
    Studies of ETGs at high redshift are expected to be less affected by the age-metallicity degeneracy because age is constrained within a narrow range of values allowed by the young age of the Universe. 
    
    Only a few stellar age and metallicity gradient measurements have been made at high redshift so far, and they are almost all based on photometric colours \citep{guo2011color,gargiulo2012spatially}. Colour-based measurements cannot break the age-metallicity degeneracy, so that they need to assume a fixed age to derive metallicity gradients or a fixed metallicity to derive possible age gradients. 
    As an exception, \cite{jafariyazani2020resolved} found a clear negative metallicity
gradient in spectroscopic data of a lensed galaxy at $z = 1.98$ that was sufficient magnified to be spatially resolved by ground-based spectroscopy. The metallicity gradient was comparable to that of local ETGs. The authors did not detect an age gradient, however.
    
    The scarcity of gradient measurements in ETGs at high redshift  is due to the current instrumental limits. Galaxies at $z \sim 2$ have an apparent size of about 0.5 arcsecond and are extremely faint.  Data at a spatial resolution of about 0.1" cannot be easily obtained with ground-based telescopes.
    Moreover, many spectral features that would be useful for measuring the age and metallicity of stellar populations at $z \sim 2$ are redshifted to infrared wavelengths. These wavelengths are strongly contaminated by our atmosphere.
    
    In the past decade, the capabilities of the Hubble Space Telescope (HST) have enabled us to overcome some of these limitations. The HST and its Wide Field Camera $3$ (WFC3) provide both photometric and spectroscopic observations with an effective spatial resolution of $\sim 0.13$ arcseconds.
    
    We present a study based on deep slitless grism spectroscopy obtained with WFC3 using the G$141$ grism. We perform a spatially resolved investigation of the age and metallicity of stellar populations in a small sample of $z \sim 2$ galaxies for the first time. 
    
    The plan of the paper is as follows. In Sect.~\ref{sec:data} we introduce our sample and the spectroscopic and photometric data. In Sect.~\ref{sec:analysis} we describe the analysis we carried out to obtain a robust measurement of age and metallicity gradients. In Sect.~\ref{sec:results} we present our results. In Sect.~\ref{sec:discussion} we discuss the constraints on the age and metallicity gradients and compare them to the literature. In Sect.~\ref{sec:conclusions} we present our conclusions and describe possible future works and implications.
    Throughout, we adopt  a standard $\Lambda$CDM cosmology with $\Omega_M = 0.3$, $\Omega_\Lambda = 0.7,$ and $H_0 = 70$ km s$^{-1}$ Mpc$^{-1}$. Magnitudes are in the AB system. 

\section{Data and sample selection}
\label{sec:data}

    The main data we used are slitless spectroscopic data obtained with the G$141$ grism and photometric data obtained with the F$160$W filter of the WFC$3$ mounted at the HST. The target of the observations is the cluster JKCS\,$041$ at $z \sim 1.8$. Spectroscopic data were taken during three separate visits with three different orientations for a total integration time of $17$ ks. These are among the deepest G$141$ grism observations taken so far, and they target a field with the largest known number of high-redshift passive galaxies. The G$141$ grism covers the wavelength range from $10750\ \AA$ to $17000 \ \AA$ at a nominal spectral resolution of $R = 130$ with a dispersion of $46.5 \ \AA$/pix. Photometric data consist of $4.5$ ks deep images of the cluster field of view in three different orientations, with a pixel scale of $0.13$ arcsec/pix.
    
    Available data are reduced and background-subtracted $2$D G$141$ spectra and calibrated background-subtracted F$160$W images for each visit \citep[see details in][]{newman2014spectroscopic}.  We extracted $1$D spectra corresponding to different portions of galaxies from the reduced spectroscopic 2D
data, thus obtaining three spectra for each visit and each galaxy: one of the inner zone, and two of the outer zone (on either side of the centre).
    
\begin{figure}
   \centering
   \includegraphics[width=0.5\textwidth]{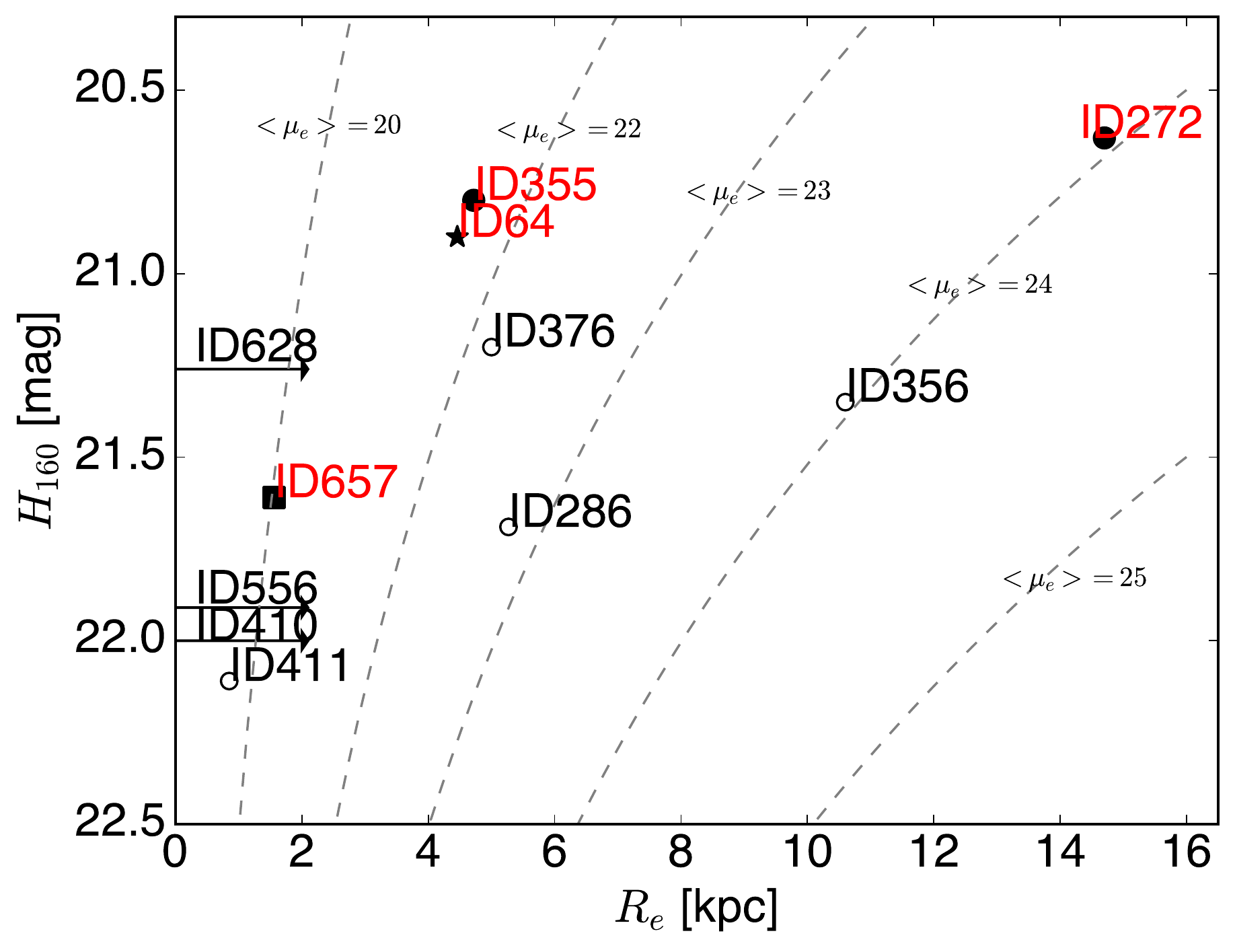}
      \caption{H$_{160}$ magnitudes vs effective radius of spectroscopic passive galaxies in the JKCS\,$041$ field of view. Dashed lines indicate the locus of points with identical surface brightness within the effective radius in mag/arcsec$^2$ (assuming the redshift of the cluster). The filled symbols indicate galaxies with constraints on metallicity and age gradients described in the text. The arrows indicate galaxies with an unknown effective radius. 
              }
         \label{fig:magn}
   \end{figure}

    The parent sample consists of all the $11$ spectroscopically passive galaxies at $1.6<z<2.4$ with $H_{160}<22.0$ in the field of view of the cluster. The targets were classified by \cite{newman2014spectroscopic} on the basis of their integrated HST G141 spectra. $\text{Seven}$ of the $11$ selected galaxies belong to the $z \sim 1.8 $ rich cluster JKCS\,$041,$ while 2 of the other 4 are foreground ($z \sim 1.6$) and 2 are background ($z \sim 2.4$) galaxies. Figure~\ref{fig:magn} shows the F$160$W magnitude of the selected targets as a function of their effective radius. Magnitudes and effective radii were calculated by \cite{newman2014spectroscopic}, except for ID$64$, for which they were calculated by us \citep[as detailed in][]{andreon2016size}. The effective surface brightnesses of the sample cover a wide range from $20$ (for compact galaxies) to $24$ (for less compact galaxies) mag/arcsec$^2$. Moreover, the effective radii of $9$  out of $11$ galaxies are smaller than one arcsecond.
    
    We discarded three galaxies (ID$411$, ID$286,$ and ID$376$) from the sample a priori because their spectra were strongly contaminated or part of the spectra lay outside the detector limit.
    In the following section, we present the analysis of the remaining $\text{eight}$ galaxies. As we explain in Sect.~\ref{sec:results}, we obtained reliable measurements of the age and metallicity parameters in different spatial zones for four out of $\text{eight}$ galaxies.
    
\noindent

%

\section{Analysis}
\label{sec:analysis}

The main goal of our analysis is to constrain the ages and metallicities of the stellar populations in different zones of the considered galaxies. In this section, we compare synthetic templates and observed spectra in a small spectral window, extracted from different zones of the galaxies, and subsequently determine the corresponding stellar population parameters. We limit the following analysis to a very narrow spectral range in order to reduce the influence of the continuum on the determination of age and metallicity parameters. The shape of the spectral continuum is affected by $\lambda$-dependent dust extinction, and this can affect the age and metallicity estimates. Practically, the selected spectral ranges correspond to $\sim 4400-5300$ \AA \ rest frame for the foreground and cluster galaxies and to $ \sim 3900-4500$ \AA \ rest frame for the background galaxies. These ranges contain spectral features that are sensitive to age and metallicity parameters in both cases.
\subsection{Stellar population models}

The galaxies in our sample were selected to be spectroscopically passive, and we adopted simple stellar population (SSP) models to represent their SFH with the aim of deriving the mean age and metallicity of their stellar population.
Our reference analysis adopted the spectrophotometric SSP models by \cite{bruzual2003stellar} combined with the \cite{chabrier2003galactic} IMF to build a synthetic spectral library. The BC$03$ models are based on the Padova tracks and isochrones combined with the MILES \citep{sanchez2007spatially} empirical stellar library. These synthetic spectra cover the range in age from $0.4$ Gyr to the lookback time between the galaxy redshift and $z_f = 6$, and in metallicity from $0.02 \ Z_\odot$ to $2.5 \ Z_\odot$.  We considered SSP because previous measurements in \cite{newman2014spectroscopic} performed on our sample showed a best-fit timescale of exponentially declining SFHs of $<0.1$ Gyr, as expected given the age of the Universe at high redshift and the passive nature of the sample. Any extended SFH therefore has by necessity a very short star formation timescale, which in turn makes them indistinguishable from data of our quality from SSP with slightly different ages \citep{longhetti2005dating}.
Nevertheless, we also considered other templates with different parameters and extended SFHs to demonstrate the robustness of our measurements (see Sect.~\ref{sec:discussion}).

%
   \begin{figure*}
   \centering
   \sidecaption
   \includegraphics[width=12cm]{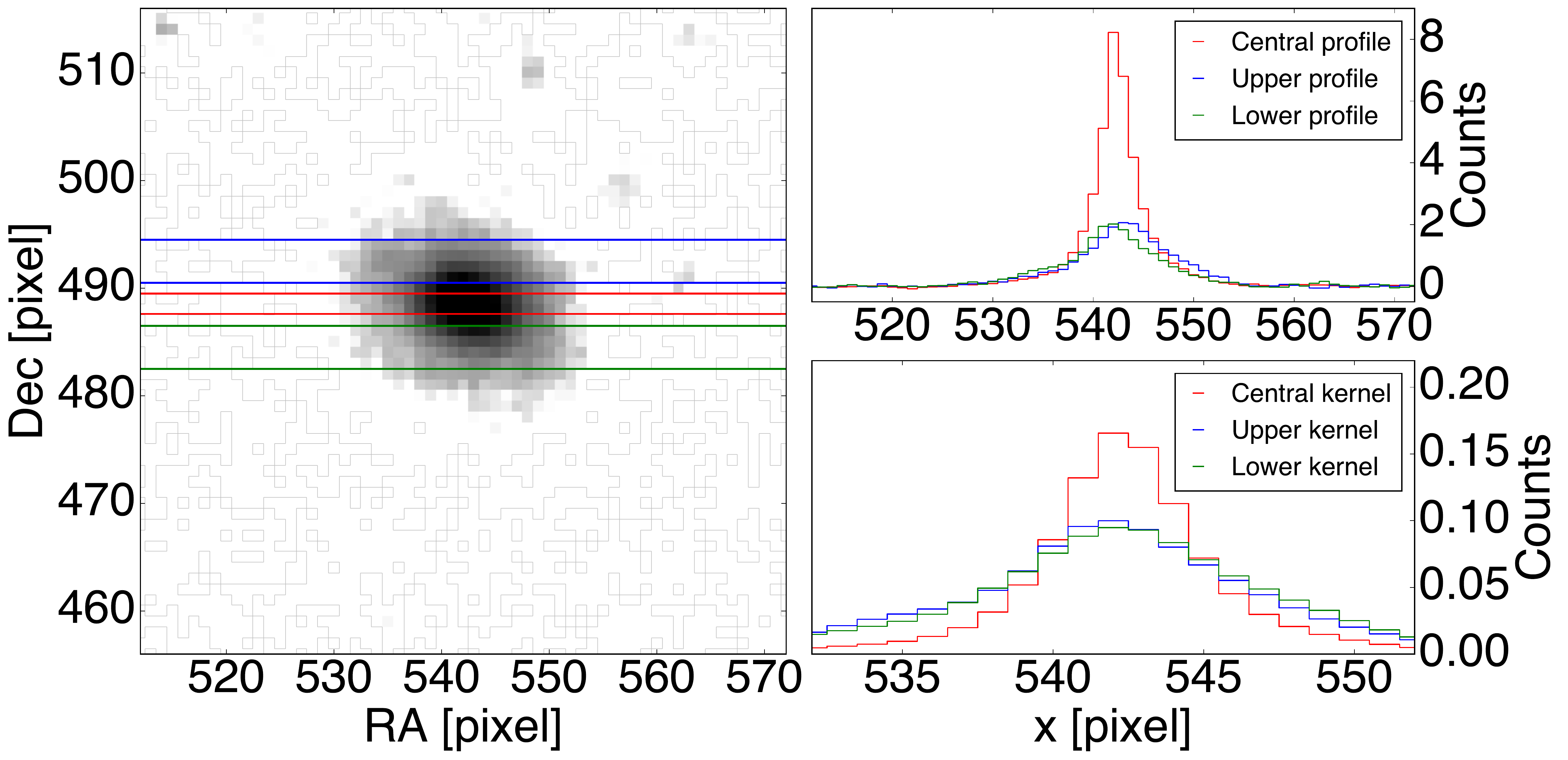}
      \caption{Example of the process to obtain the convolution kernel for a single visit of the galaxy ID$355$. Left panel: F$160$W image of ID$355$ (single visit). Upper right panel: Light profile of ID$355$. Each pixel along the dispersion side is $0.121$ arcseconds wide, corresponding to $46.5 \ \AA$. Lower right panel: Convolution kernel of ID$355$.
              }
         \label{fig:improker}
   \end{figure*}
%
%
\subsection{Degradation of the synthetic spectra}
\label{sub:degrade}
Before they are compared with observations, synthetic spectra need to be: redshifted to the galaxy reference frame, degraded to the spectral resolution of the observed spectra (the spectral resolution is given by the combination of the nominal resolution of the instrument with the shape of the light profile along the dimension of the spectrum extraction), and rebinned to the pixel scale of the observed spectra.
For numerical convenience (the galaxy radial profile is poorly known below the pixel resolution), we inverted the order of the last two processes in the analysis. We tested that this choice did not affect the final result by considering the worst-case scenario, which is an unresolved source with a Gaussian spectral feature comparable to the resolution of the observed spectra.
For the second process, the kernel that was used to degrade the theoretical templates was derived by convolving the galaxy light profile with the nominal resolution of the instrument. 

It is highly important to use the light profile of the galaxy to build the kernel because different galaxy profiles cause a different effective spectral resolution. The correlation with morphology means that the spectral resolution decreases with increasing galaxy size \citep[morphological broadening;][]{van2011first}.
At the resolution of our data, the broadening caused by the galaxy extent is the dominant effect, corresponding to $\sim 2000$ km/s rest-frame, and the dynamical motions within the galaxy can be ignored. 

In order to measure the stellar population properties in different zones within the galaxies and to derive their gradients, we divided each galaxy into three zones and separately extracted the corresponding spectra from  each G$141$ visit. We used the F$160$W image to extract the light profile of the galaxies corresponding to each of the extracted spectra because it matches the spectral coverage of the G$141$ grism. The extraction zones are larger in the outer than in the inner zone in order to partially compensate for the fact that the signal-to-noise ratio (S/N) is lower in the outskirts of the galaxies than in the central zones.

For each of the extraction zones, we derived the corresponding kernel function by convolving the light profile of the galaxy with a Gaussian with $\sigma=46.5 \ \AA,$ representing the instrumental resolution of G$141$ grism. The resulting kernels were used to convolve the synthetic templates.

Figure~\ref{fig:improker} shows an example of the process described above. It was  used to obtain the convolution kernel for a single visit of galaxy ID$355$. The left panel shows the division of the galaxy into three zones, corresponding to the extraction zones of the spectra. The upper right panel shows the three light profiles obtained by summing the pixels along the dispersion direction. The lower right panel shows the kernels obtained by convolving the light profile with the spectral resolution of the G$141$ grism. The outer kernels are broader and flatter than the inner kernel because the selected outer zones are spatially larger and fainter than the inner zone.
Therefore the spectral features are broader in the outer zones than the inner zone.

\begin{figure}
   \centering
   \includegraphics[width=0.5\textwidth]{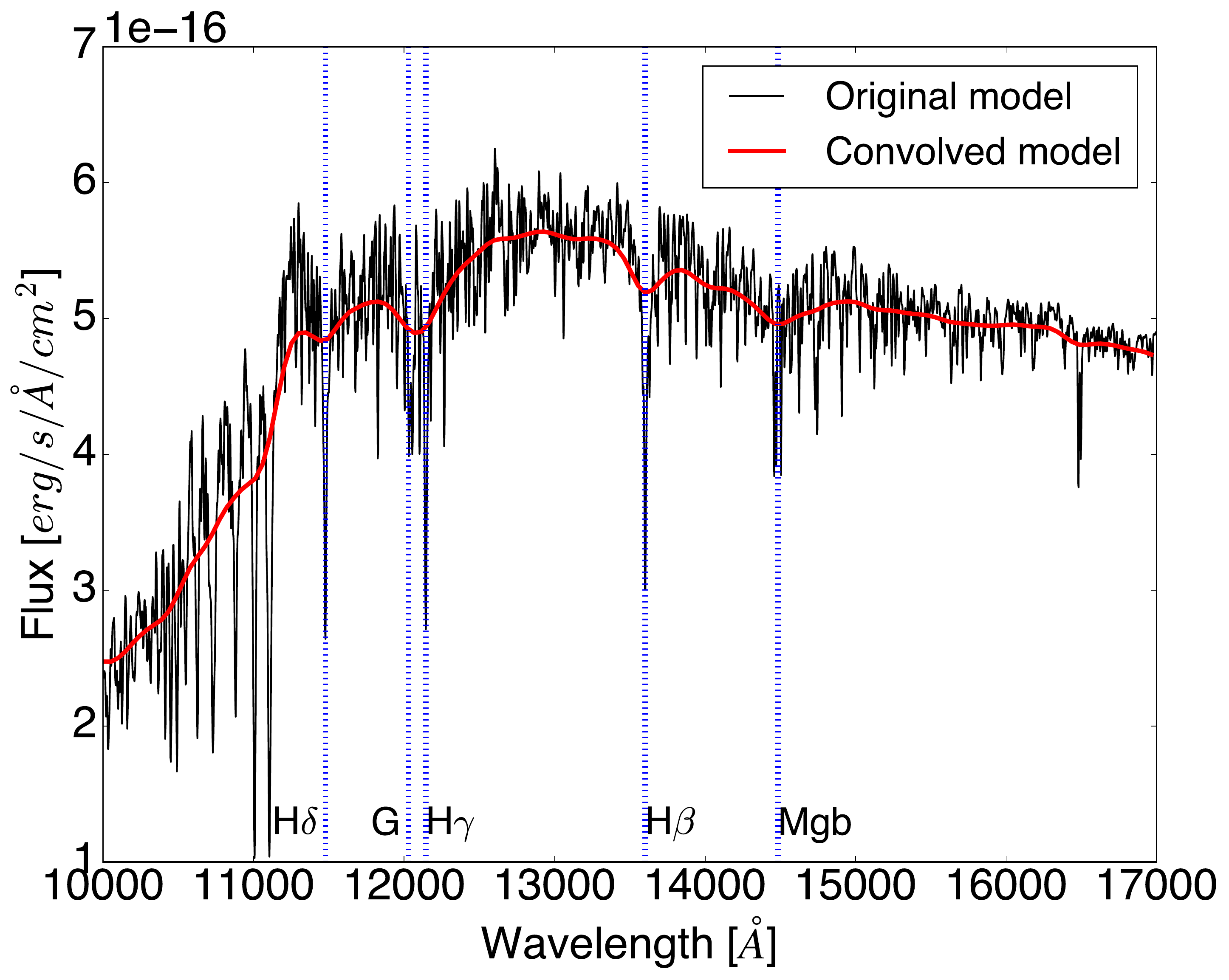}
      \caption{Rebinning and convolution effects. In this example, we used a synthetic spectrum with solar metallicity and an age of $1$ Gyr. The black line indicates the original redshifted spectrum, and the red line represents the rebinned and convolved spectrum. The vertical blue lines indicate the absorption lines we considered.
              }
         \label{fig:confr}
   \end{figure}

Figure~\ref{fig:confr} shows an example of a synthetic spectrum redshifted to the ID$355$ reference frame and the same synthetic spectrum at the spectral resolution and binning of the observations. Despite the degradation of the spectrum, the absorption lines that are sensitive to the age and metallicity of the stellar populations (H$\delta$, H$\gamma$, H$\beta,$ and Mgb) are visible, suggesting that it is possible to measure the age and metallicity parameters at this spectral resolution.

\subsection{Joint fit}
\label{sub:joint}
Our statistical analysis was conducted with a Bayesian approach, which provides a powerful framework for inferring the age and metallicity of stellar populations in galaxies.
Because the S/Ns of the observed spectra were low and diverse, we performed a joint fit of all the available data for each extraction zone in each galaxy. Upper and lower outer zones were considered together. The joint fit prevents the possible high noise of a single visit from degrading the signal coming from the other visits.

We compared the observed data with the redshifted, rebinned, and convolved synthetic templates within a narrow spectral range, and we computed the posterior probability of the age and metallicity parameters using the likelihood given by $\mathcal{L} = e^{-\chi^2/2}$, where $\chi^2$ is
\begin{equation}
    \chi^2 = \sum_{i}{\left(\frac{F_{syn_i}-F_{obs_i}}{\sigma_{obs_i}}\right)}^2
,\end{equation}
where $F_{syn}$ is the flux of the synthetic spectrum, and $F_{obs}$ is the flux of the observed spectrum with the error $\sigma_{obs}$. The index $i$ indicates the i-th pixel on which the calculation takes place. 

The total likelihood is given by the product of the likelihood of each visit. Before multiplying likelihoods, the observed spectra were inspected to determine possible contamination or cosmetic problems, such as reduction residuals or cosmic rays. The few spectra with $p$-values\footnote{The p-value is the probability of observing more extreme data under the assumption that the null hypothesis is correct.} lower than $0.002$ were discarded.

We assumed a uniform prior for age (from $0.4$ Gyr to the lookback time between the galaxy redshift and $z_f = 6$) because we considered only a small range, and a logarithmically uniform prior for metallicity (from $0.02 \ Z_\odot$ to $2.5 \ Z_\odot$), the latter following previous works \citep[e.g.][]{gallazzi2014charting,morishita2018metal,zibetti2020insights}.
We explored the parameter space with a Markov chain Monte Carlo method \citep[MCMC;][]{gilks2005m}. The 68\% probability interval is delimited by the $16^\circ$ and $84^\circ$ percentile. We also considered different prior distributions (uniform for metallicity, logarithmically uniform for age) to demonstrate the robustness of our measurements (see Sect.~\ref{sec:discussion}).

\section{Results}
\label{sec:results}

%
   \begin{figure}
   \centering
   \includegraphics[width=0.5\textwidth]{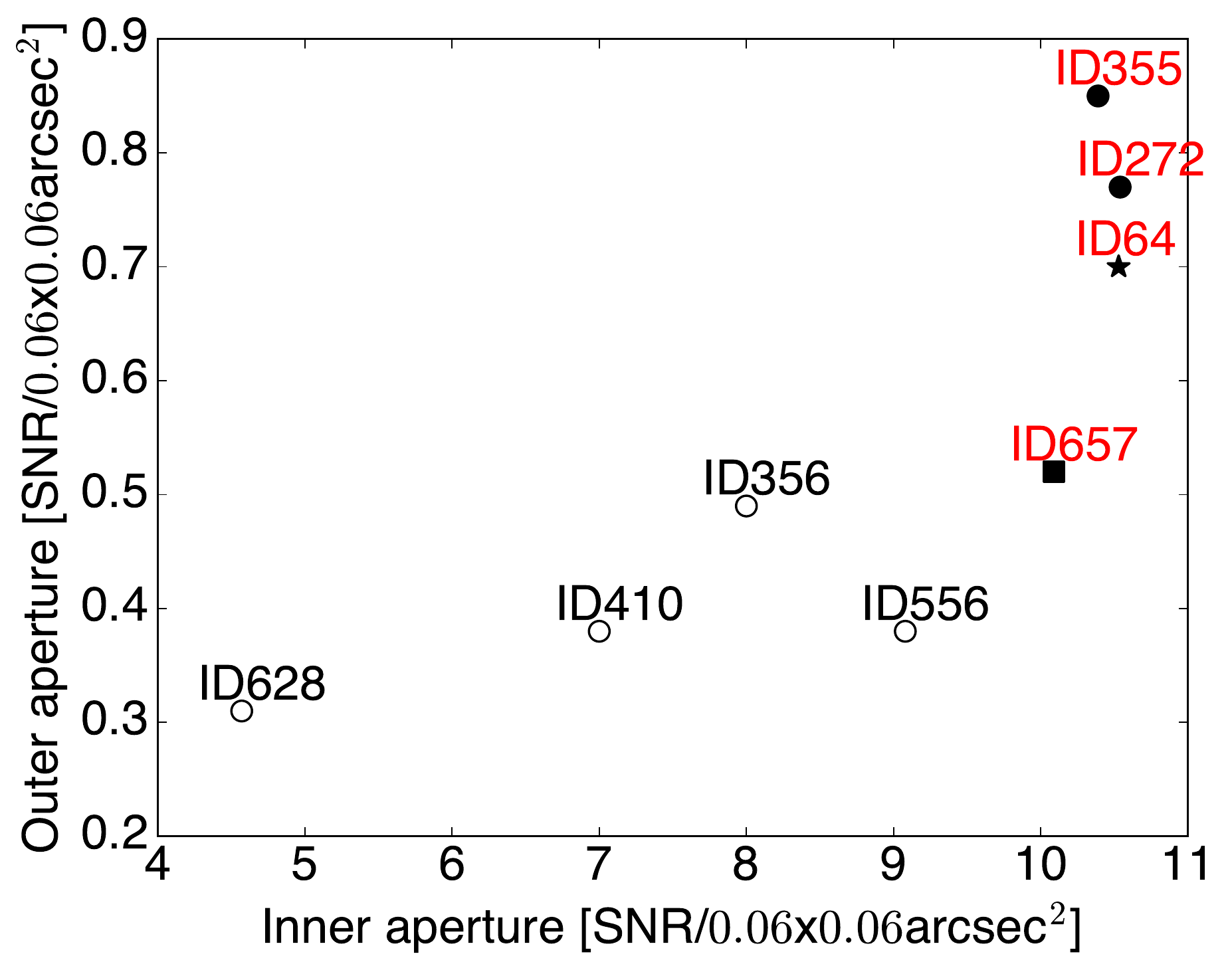}
      \caption{F$160$W S/N of the inner and outer zones. The filled symbols indicate the galaxies described in the text. S/Ns are given for a single imaging pixel ($0.06x0.06$ arcsec$^2$), but they were calculated as the average S/N over larger areas ($\sim 10$ for the inner zone, and $\sim 400$ for the outer zone).
              }
         \label{fig:snraperture}
   \end{figure}

\begin{table*}
\caption{Main characteristics of the four selected galaxies.}             
\label{table:selected}      
\centering          
\begin{tabular}{l c r c c c}     
\hline\hline       
Galaxy & H$_{160}$ & $R_e$ & $\log M_\star/M_\odot$ & Redshift & Lum. w. radius of the inner and outer zones${^{(a)}}$ \\
& [mag] & [kpc] & & & [kpc] \\
\hline                    
  ID355 & 20.80 & 4.7 & 11.52 & 1.798 & 0.4 - 3.9 \\      
   ID64 & 20.90 &  4.5${^{(a)}}$  & 11.66 & 2.415 & 0.7 - 3.4 \\
   ID272 & 20.63 & 14.7 & 11.71  & 1.798 & 0.4 - 5.2 \\
   ID657 & 21.61 & 1.6 & 11.11 & 1.812 & 0.5 - 2.6 \\
\hline                  
\end{tabular}
\tablefoot{
H$_{160}$, $R_e$, $\log M_\star/M_\odot$ and redshift are from \cite{newman2014spectroscopic}.\\
\tablefoottext{a}{Values measured in this work.}}
\end{table*}

   \begin{figure*}
   \centering
   \includegraphics[width=0.95\textwidth]{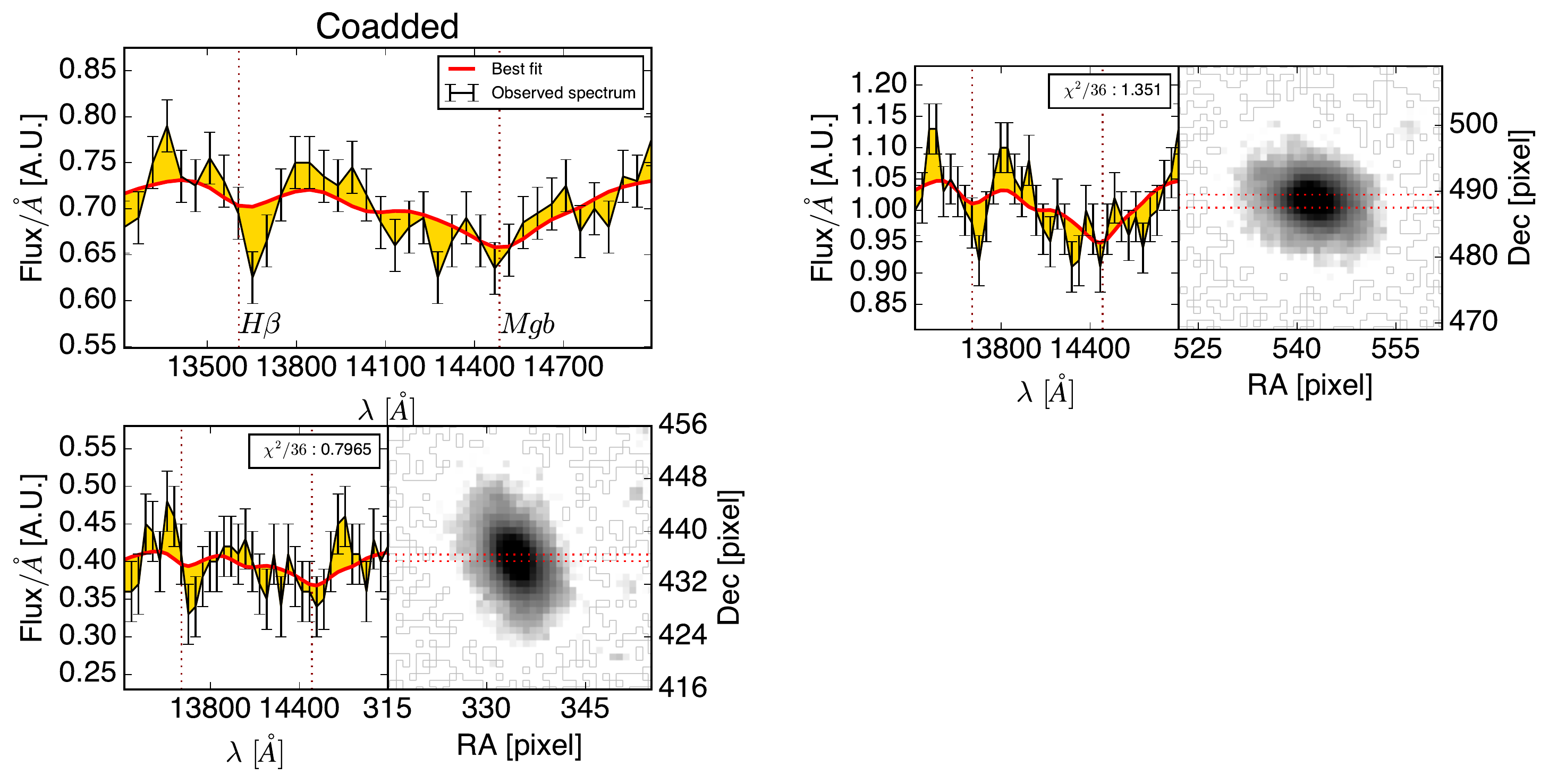}
      \caption{Extracted spectra and the corresponding best-fit templates in the inner zone of galaxy ID$355$. Upper left panel: Coadded spectrum of the inner zone of ID$355$ (line with errors) with the best-fit model in arbitrary units. Other panels: Single-visit observed spectrum of the inner zone of ID$355$ with the best-fit model and the corresponding F$160$W images with the related extraction. 
              }
         \label{fig:central}
   \end{figure*}

   \begin{figure*}
   \centering
   \includegraphics[width=0.95\textwidth]{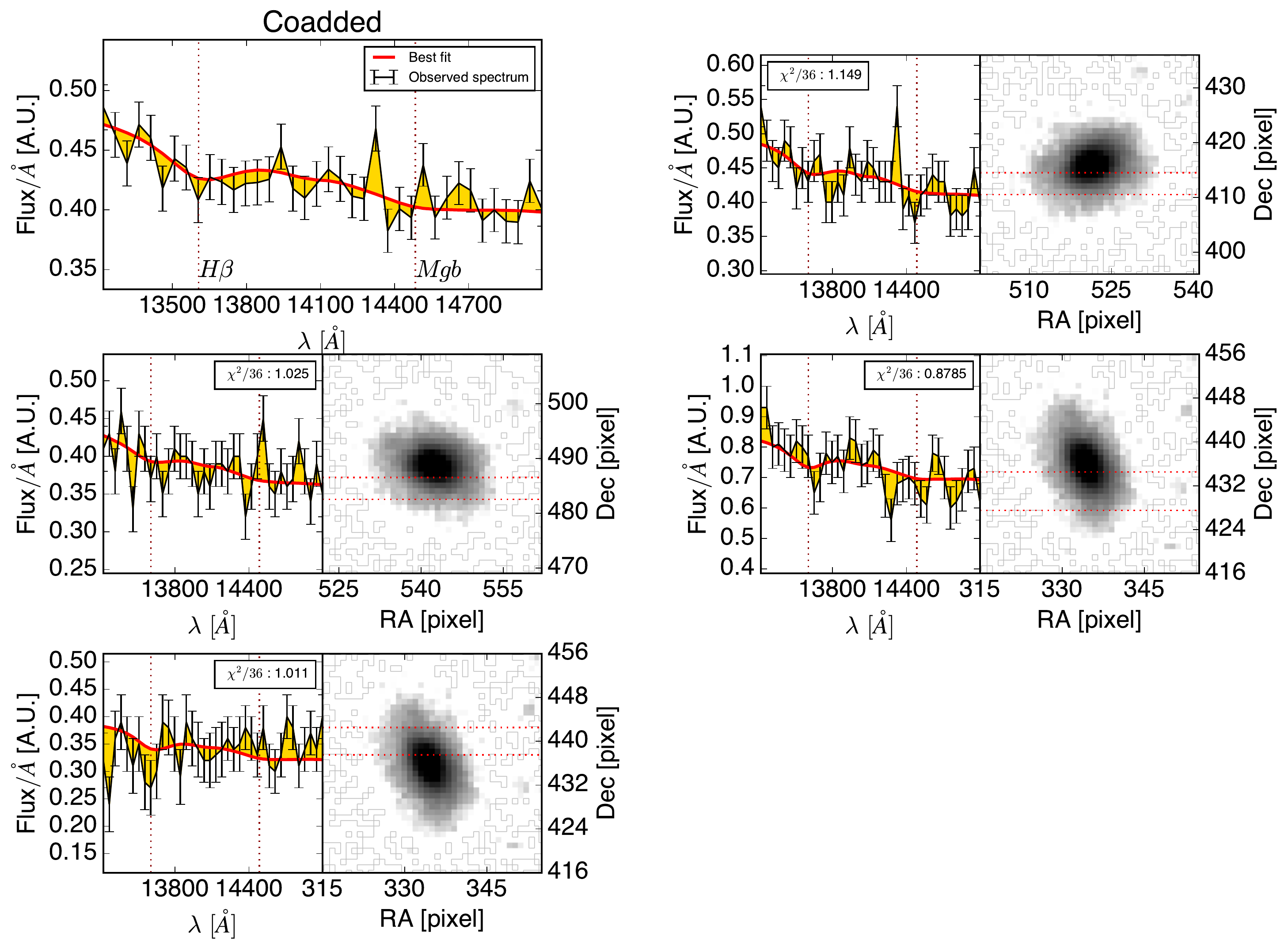}
      \caption{Same as Fig.~\ref{fig:central} for the outer zone of ID$355$.
              }
         \label{fig:ext}
   \end{figure*}

   \begin{figure}
   \centering
   \includegraphics[width=0.5\textwidth]{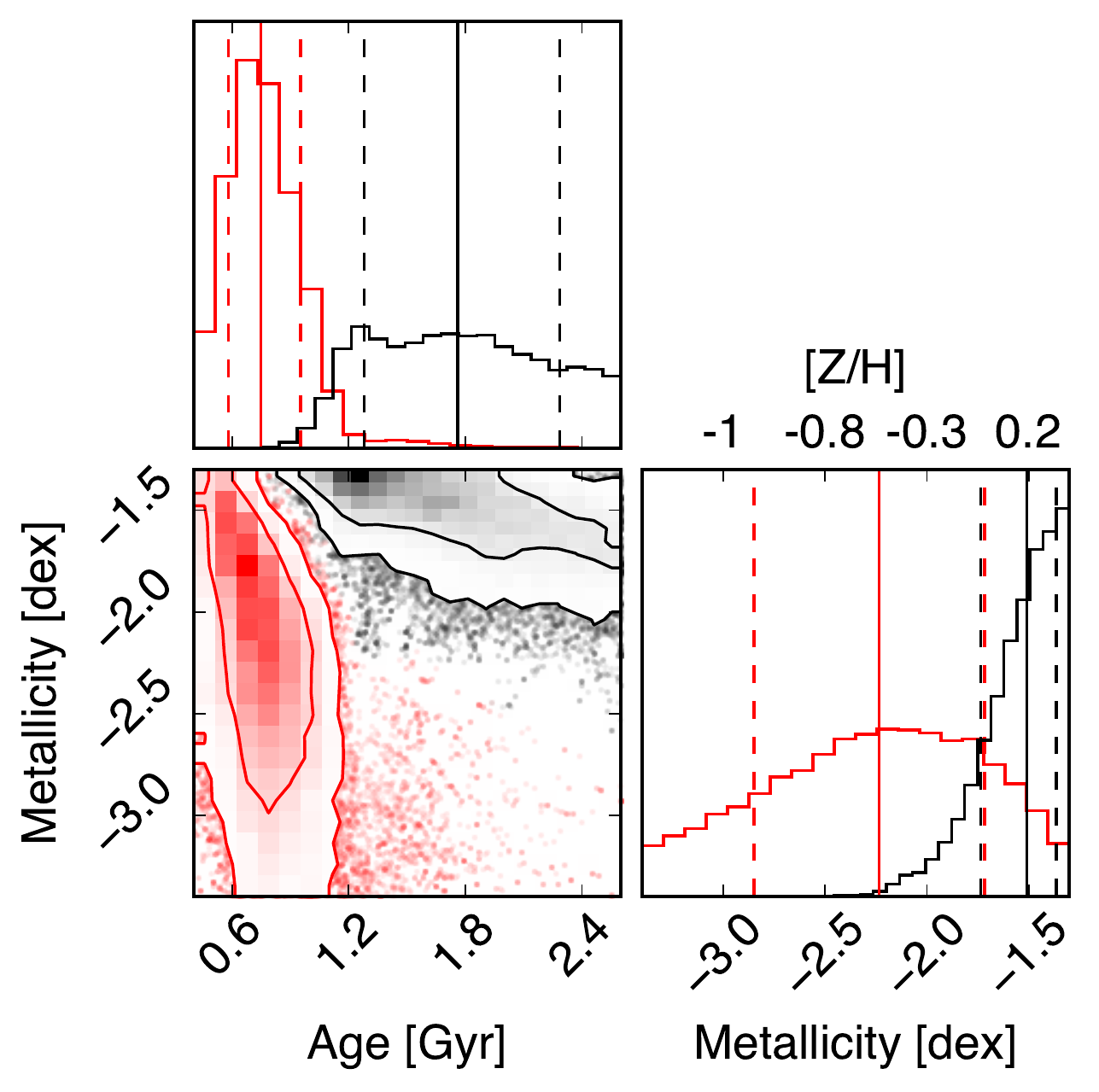}
      \caption{Combined age and metallicity distributions obtained for ID$355$. Lower left panel: Joint probability distribution of age and metallicity of the inner (black) and outer (red) zones of ID$355$.  Contours are at 68\% and 95\% probability.
      Upper left panel: Marginalised probability for the age. Lower right panel: Marginalised probability for the metallicity. Median and 16\%\ and 84\% intervals are indicated by solid and dashed lines, respectively.
              }
         \label{fig:joint}
   \end{figure}

%
   \begin{figure}
   \centering
   \includegraphics[width=0.5\textwidth]{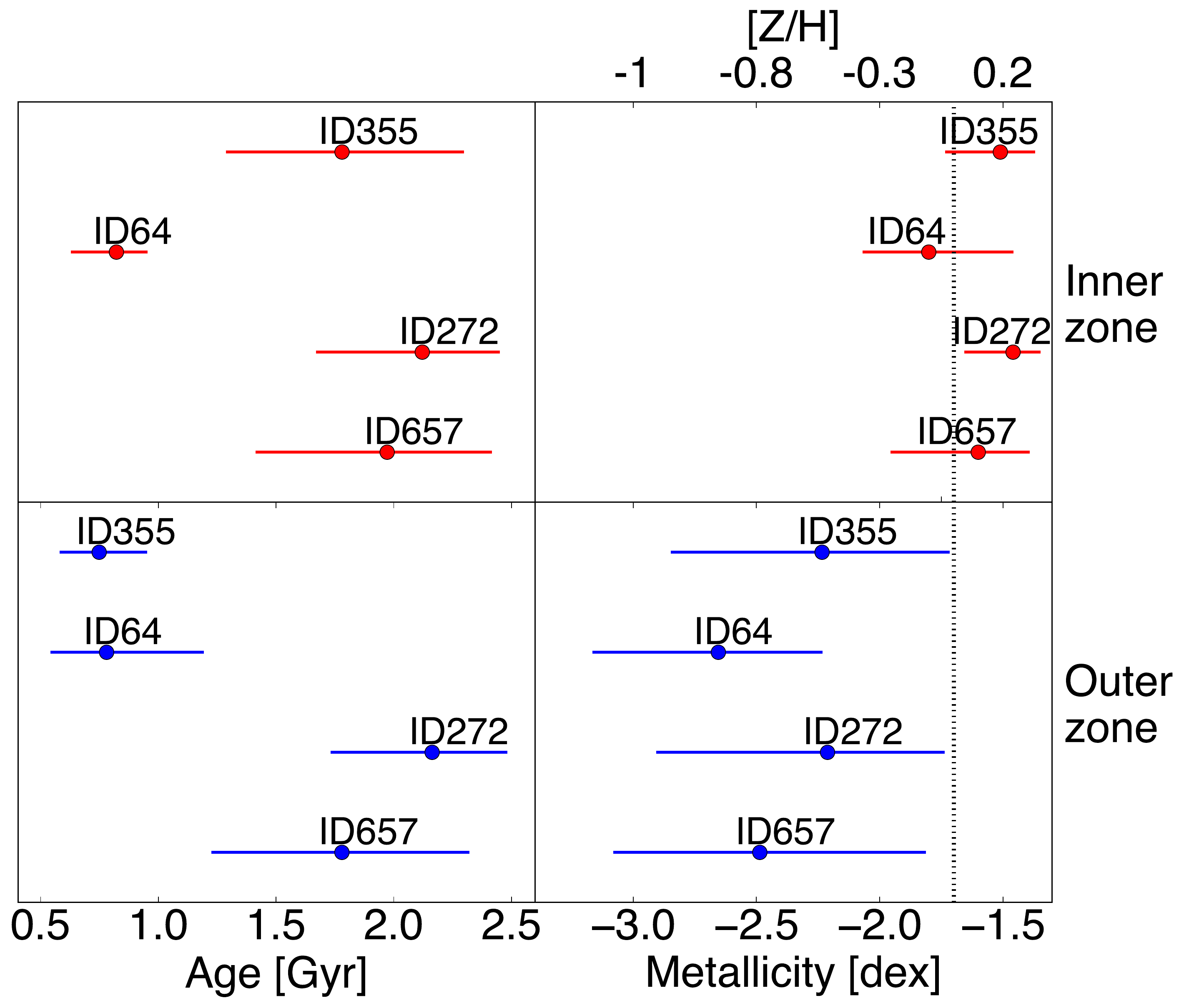}
      \caption{Age (left panel) and metallicity (right panel) median values in the inner and outer zones of the sample. Solid lines show the $1\sigma$ errors. The vertical dashed line indicates solar metallicity.
              }
         \label{fig:agemet}
   \end{figure}

   \begin{figure*}
   \centering
   \includegraphics[width = 0.8\textwidth]{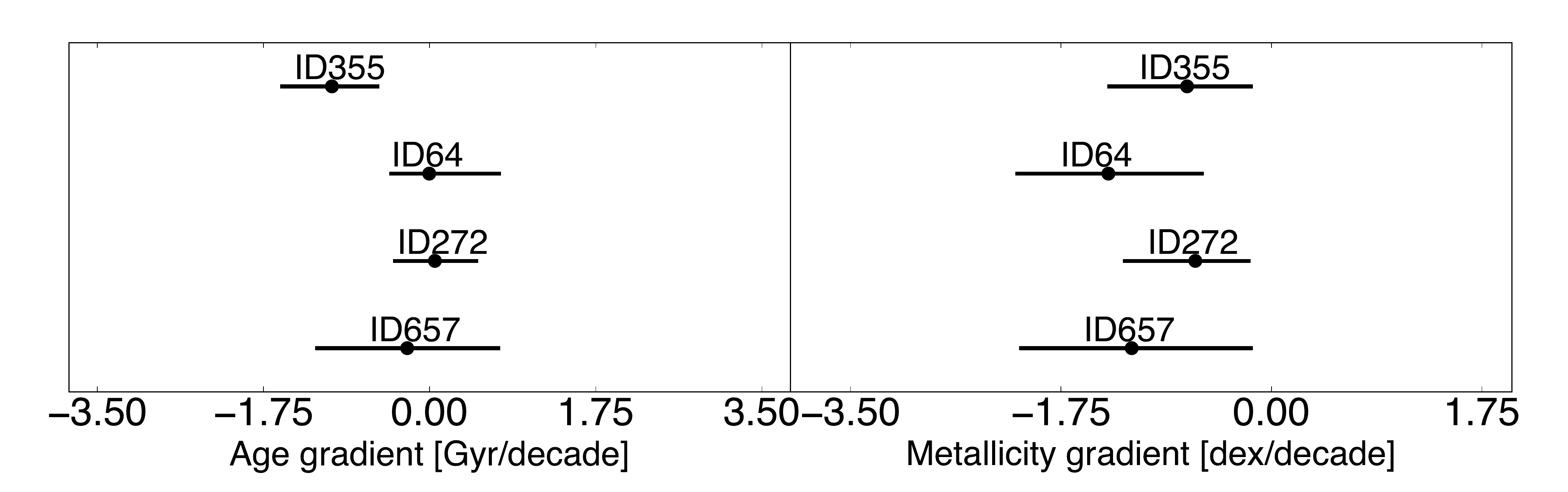}
   \caption{Age (left panel) and metallicity (right panel) gradients of the sample. Solid lines show the $1\sigma$ errors. }
              \label{fig:gradients}
    \end{figure*}

As detailed in the previous paragraph, we extracted the spectra corresponding to the inner and outer zones of the $\text{eight}$ galaxies introduced in Sect.~\ref{sec:data}. We measured the best fit and the joint probability distributions of the age and metallicity parameters in each zone for each of them.
 $\text{Sixty-eight percent}$ of the posterior in four out of the $ \text{eight}$  galaxies (ID$356$, ID$410$, ID$556$, and ID$628$) is nearly equal to $68\%$ of the assumed prior, and age and metallicity remain entirely unconstrained in at least one of the two zones (inner or outer), while for the remaining four galaxies (ID$355$, ID$272$, ID$657$, and ID$64$), the errors are smaller.
Figure~\ref{fig:snraperture} shows the S/N in the outer zone as function of the S/N in the inner zone for the combined F$160$W image of all $\text{eight}$ galaxies, and it suggests a way to preselect galaxies that are more likely to return useful constraints from a spectroscopic analysis without actually performing it. We calculated the photometric S/N using an aperture with a radius of two imaging pixels for the inner zone and a typical annulus with an inner radius of 4 pixels and an outer radius of 12 pixels for the outer zone. This matches the typical spectral apertures. 
Three galaxies (ID$355$, ID$272,$ and ID$64$) are different from the remaining sample, in particular, they have a higher S/N in the outer zone. Galaxy ID$657$ represents a borderline case. It has one of the highest S/Ns in the inner zone and the best S/N in the outer zone of the other five galaxies. These four galaxies are characterised by the highest S/N in the outer regions (filled symbols) and are those for which our analysis obtained the smallest errors when we estimated age and metallicity in the two zones.
Galaxies reported with empty circles represent galaxies for which the resulting age and metallicity estimates are affected by large errors, in particular in the outer zone, comparable to the prior we considered. For these galaxies, we cannot obtain information about their possible gradients even if it is present.
The four galaxies for which we obtained the smallest errors therefore correspond to the galaxies with the highest S/N in the outer zone. 
This criterion could be used to find a priori suitable candidates in future surveys to study age and metallicity gradients without the need of performing a full spectroscopic analysis. Differences in the depths between the considered data need to be accounted for.
    
In the following, we present our results for the four galaxies with the smallest errors (ID$355$, ID$272$, ID$657$, and ID$64$). They are also presented in Table~\ref{table:selected}. Three galaxies are members of JKCS\,$041$, at $z \sim 1.8$, and their fitted rest-frame wavelength range includes H$\beta$ and Mgb features, which are sensitive to the age and metallicity of galaxy stellar populations, respectively. For the higher-redshift galaxy ID$64$, we considered the rest-frame range that includes H$\delta$ and H$\gamma$ lines (sensitive to the stellar age) and G band (4300) (which is also sensitive to stellar metallicity).

Figures~\ref{fig:central} and~\ref{fig:ext} show the extracted spectra of galaxy ID$355$  and the corresponding best-fit templates in the inner and outer zone, respectively. Figure~\ref{fig:joint} shows the combined age and metallicity distributions obtained for ID$355$. Figures of the results obtained from the analysis of the other galaxies are shown in Appendix A.

Figure~\ref{fig:agemet} summarises our results for the four galaxies. It reports the median value of age and metallicity for each zone and their 68\% probability intervals. We find solar and supersolar metallicities in the inner zones of all the galaxies and subsolar values in the outer zones, indicating  negative metallicity gradients in all four galaxies. Three out of the four galaxies the inner and outer zone have comparable ages, while ID$355$ presents an older stellar population in the inner zone than in the outskirts.

Finally, Fig.~\ref{fig:gradients} shows the calculated median values of the age and metallicity gradients for the four galaxies. The gradients are defined as
\begin{equation}
    \nabla \log Z = \frac{\Delta \log Z}{\Delta \log R} \ \ \  \nabla t = \frac{\Delta t}{\Delta \log R}
,\end{equation}
where $\log$ is the base-$10$ logarithm, $Z$ is the metallicity, $t$ is the age (Gyr), and $\Delta \log R$ is the difference between the luminosity-weighted radius of the inner and outer extraction zones. Median values and errors are derived from the Bayesian probability density fuction of the gradients.
Table~\ref{table:valerr} shows the median values of the marginalised posterior distribution of age and metallicity in the inner zone and their gradients.
All four galaxies present negative metallicity gradients, even if their significance is only slightly above $1\sigma$. Only one of the four galaxies, namely ID$355$, displays a negative age gradient. We obtained an average error of $0.7$ dex/decade for the metallicity gradients and $0.6$ Gyr/decade for the age gradients.

Our gradient definition does not account for point spread function (PSF) effects, therefore the derived gradients underestimate the real gradients of the galaxies and should be handled with caution when compared with ground-based measurements with different (lower) spatial resolution. Moreover, the contamination of the spectra of the inner regions by the outer regions that lie along the dispersion direction might partially wash out the real gradients of the galaxies.

\begin{table*}
\caption{Median values of the marginalised posterior distribution of age and metallicity in the inner zone and their gradients. The errors of the median values refer to the 16th$^{}$ and 84th$^{}$ percentile.}             
\label{table:valerr}      
\centering          
\begin{tabular}{l c c c c c}     
\hline\hline       
Galaxy & Age$_{in}$ & Z$_{in}$ & Z$_{in}$ & $\nabla$ Age & $\nabla$ Z\\
& [Gyr] & [dex] & [Z/H] & [Gyr/decade] & [dex/decade] \\
\hline   \\                  
  ID355 & 1.78$^{+0.51}_{-0.49}$ & -1.51$^{+0.14}_{-0.22}$ & 0.19$^{+0.14}_{-0.22}$ & -1.03$^{+0.50}_{-0.55}$ & -1.70$^{+0.55}_{-0.66}$ \\\\      
   ID64 & 0.82$^{+0.13}_{-0.19}$ & -1.80$^{+0.34}_{-0.27}$ & -0.1$^{+0.34}_{-0.27}$ & -0.00$^{+0.76}_{-0.42}$ & -1.36$^{+0.79}_{-0.78}$\\\\
   ID272 & 2.12$^{+0.33}_{-0.45}$ & -1.46$^{+0.11}_{-0.20}$ & 0.24$^{+0.11}_{-0.20}$ & 0.06$^{+0.45}_{-0.44}$ & -0.63$^{+0.46}_{-0.60}$\\\\
   ID657 & 1.97$^{+0.44}_{-0.56}$ & -1.60$^{+0.21}_{-0.36}$ & 0.1$^{+0.21}_{-0.36}$ & -0.24$^{+0.98}_{-0.97}$ & -1.16$^{+1.00}_{-0.94}$\\\\
\hline                  
\end{tabular}
\end{table*}
\section{Discussion}
\label{sec:discussion}

%
\begin{table*}
\caption{Assumptions used for the sensitivity analysis.}             
\label{table:sensitivity}      
\centering                          
\begin{tabular}{l l l l l l l}        
\hline\hline                 
Model & Stellar lib. & IMF & Max $z_f$ & Z/Z$_\odot$ & Prior shapes & SFH\\    
&&&&& (metallicity-age)&\\
\hline                        
   BC03 & MILES & Chabrier & 6 & 0.02 - 2.5 & log-lin & SSP \\      
   BC03 & MILES & Chabrier & 11 & 0.02 - 2.5 & log-lin & SSP  \\
   BC03 & MILES & Chabrier & 6 & 0.02 - 2.5 & lin-lin & SSP \\ 
   BC03 & MILES & Chabrier & 6 & 0.02 - 2.5 & log-log & SSP \\ 
   BC03 & MILES & Chabrier & 6 & 0.02 - 2.5 & lin-lin & Top-hat 0.5 Gyr \\ 
   BC03 & STELIB & Chabrier & 6 & 0.02 - 2.5 & log-lin & SSP  \\
   Maraston & MILES & Kroupa & 6 & 0.02 - 2.5 & log-lin & SSP \\
   BC03 & STELIB & -1.5 slope & 6 & 0.2 - 2.5 & log-lin & SSP \\
   BC03 & STELIB & Salpeter & 6 & 0.2 - 2.5 & log-lin & SSP \\
   BC03 & STELIB & -3.5 slope & 6 & 0.2 - 2.5 & log-lin & SSP  \\
   
\hline                                   
\end{tabular}
\end{table*}
%

   \begin{figure*}
   \centering
   \includegraphics[width=0.9\textwidth]{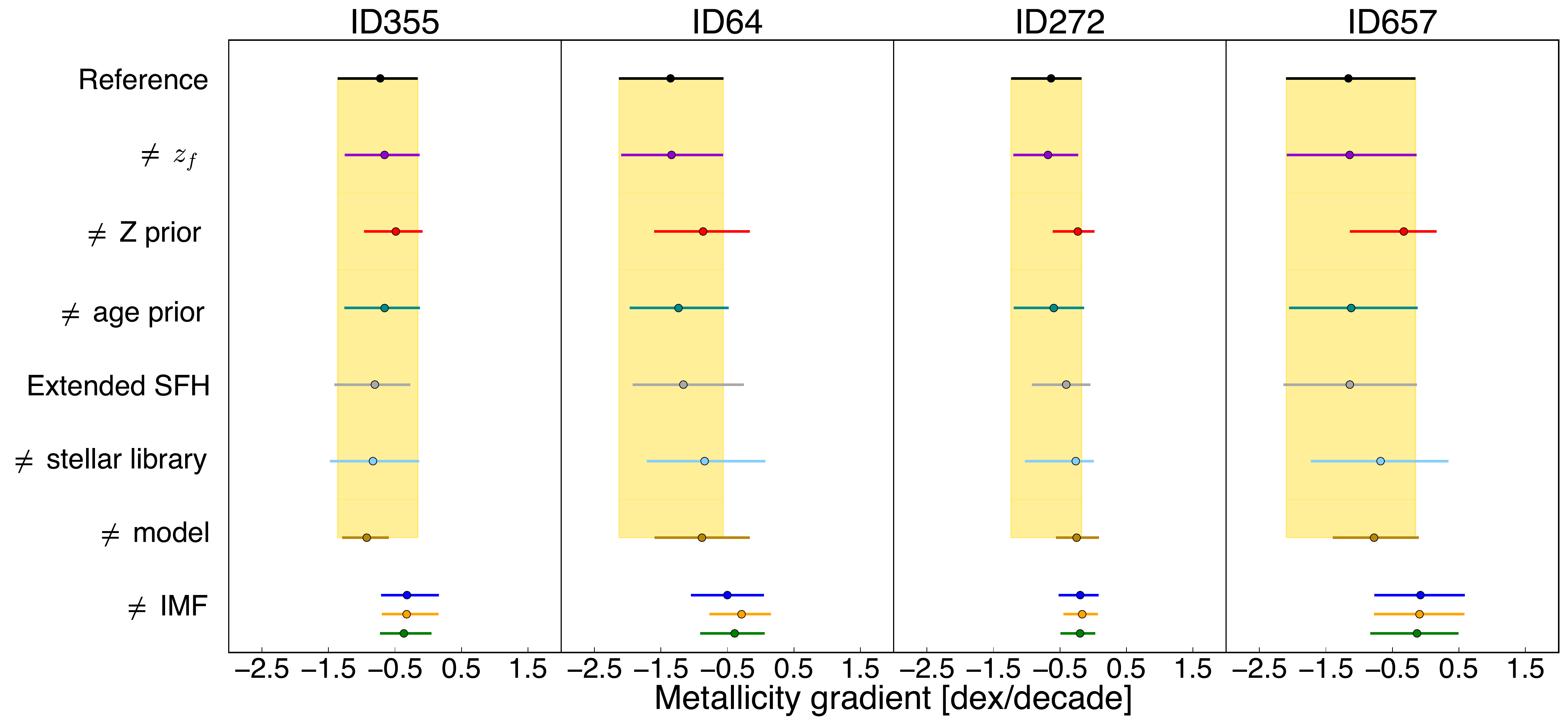}
      \caption{Sensitivity analysis for the metallicity gradients. The reference analysis (black) adopts BC03 models, the MILES stellar library, a Chabrier IMF, and $z_f = 6$. The result we obtained when we adopted a different $z_f = 11$ is reported in violet, a different metallicity prior (uniform) in red, a different age prior (logaritmically uniform) in cyan, an extended star formation timescale ($0.5$ Gyr) in grey, a different stellar library (STELIB) in light blue, and a different model (Maraston) in brown.
      The blue, yellow, and green symbols are gradients derived by adopting a different IMF ($-1.5$ slope, Salpeter, and $-3.5$ slope, respectively). These slopes are systematically flatter because the range of metallicities considered in this comparison is smaller.
              }
         \label{fig:sensmet}
   \end{figure*}
   \begin{figure*}
   \centering

   \includegraphics[width=0.9\textwidth]{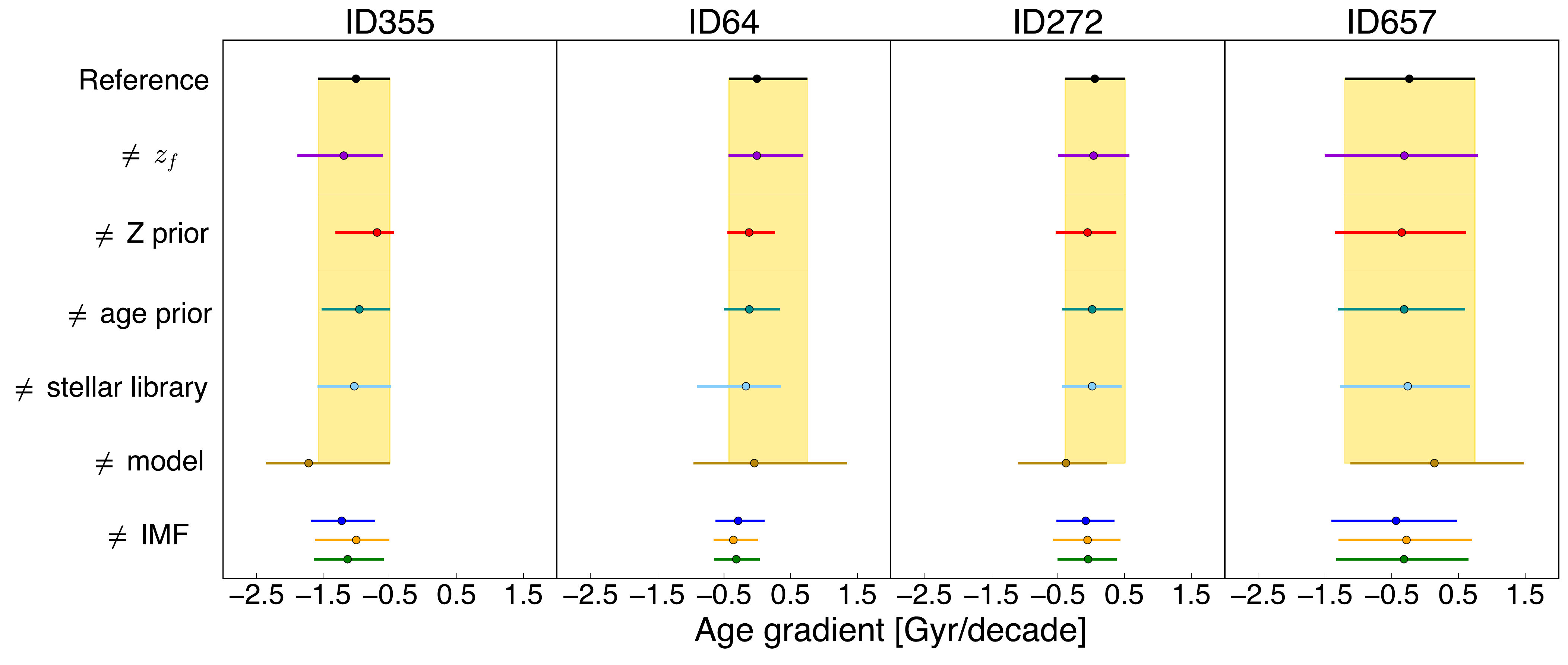}
      \caption{Sensitivity analysis for the age gradients, colour-coded as in Fig.\ref{fig:sensmet}.
              }
         \label{fig:sens}
   \end{figure*}

   \begin{figure*}
   \centering
   \includegraphics[width = 0.8\textwidth]{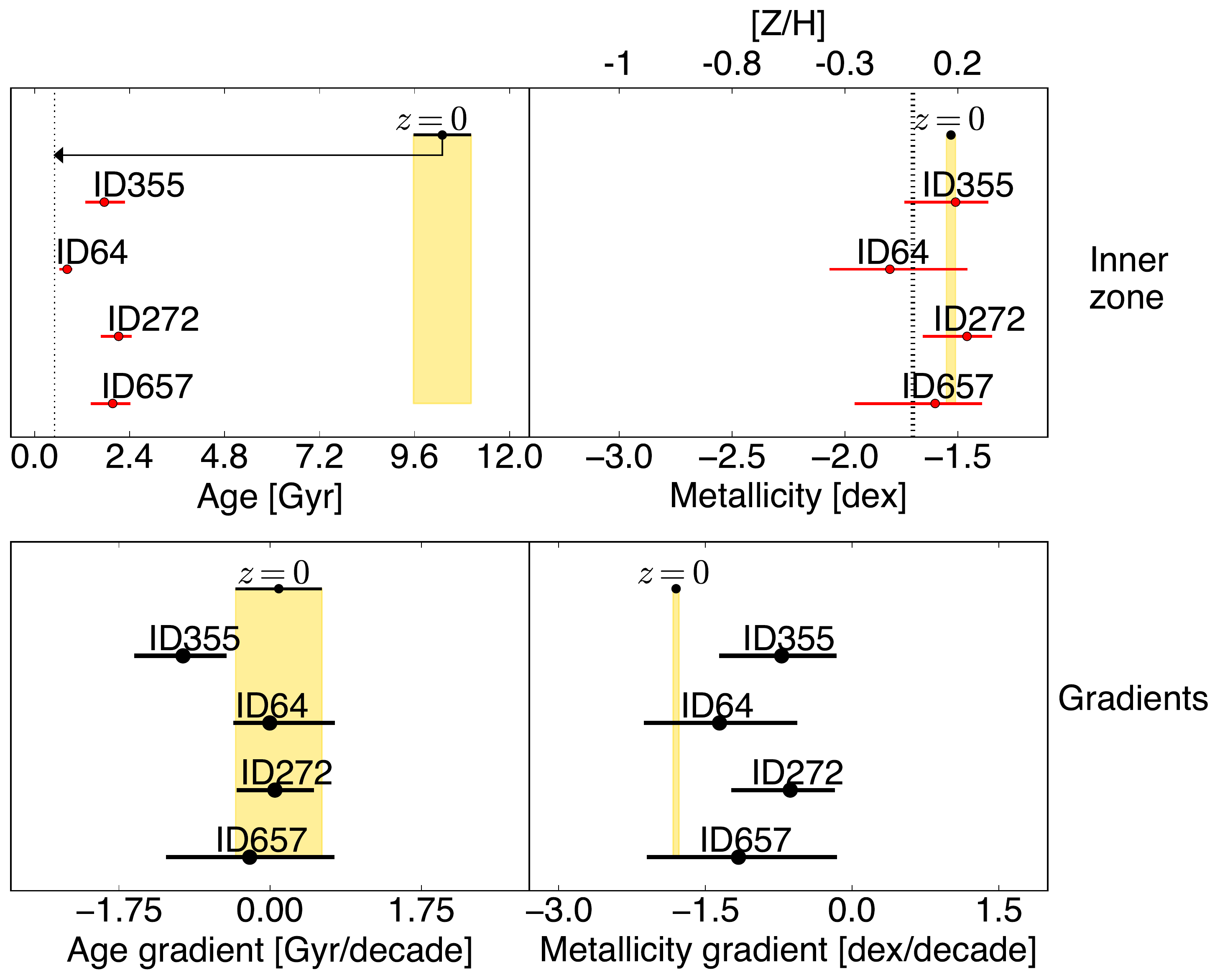}

   \caption{Comparison between results for the local Universe \citep{mehlert2003spatially} and our stellar age and metallicity parameters in the inner zone and their gradients. The arrow in the upper left panel indicates the lookback time between $z = 0$ and $z = 1.8$. The vertical dashed line indicates solar metallicity.}
              \label{fig:agemetgrad}
    \end{figure*}

We obtained reliable measurements of age and metallicity parameters in different spatial zones of four massive and spectroscopically passive galaxies. The entire sample exhibits negative metallicity gradients. 
Because of the very low spectral resolution combined with the low S/N that characterise our data, the $68\%$ of the age posteriors in our analysis are of the order of $68\%$ of the assumed prior (i.e. $16-84\%$ percentile range). Our ability to detect possible age gradients in the galaxies exists only for age differences larger than the average value of $1.2$ Gyr (see Fig.~\ref{fig:gradients}), to be compared with the mean age of their stellar content, $\approx 2$ Gyr. A gradient like this has only been revealed in one of the four sample galaxies (ID$355$), while the other galaxies are consistent with a constant age with radius.

Before discussing the results, we verified whether they depend on the assumed model templates,  on their selected parameters, or on the assumed priors. Moreover, in Appendix A, we show the robustness of our age and metallicity estimates by adding the effect of dust extinction as a further fitting parameter.

Figures~\ref{fig:sensmet} and~\ref{fig:sens} show the metallicity and age gradients, respectively, obtained in the reference case (i.e. based on the  standard assumptions discussed in Sect.~\ref{sec:analysis}) and in test cases based on different assumptions listed in Table~\ref{table:sensitivity}.
The metallicity gradient estimates are very stable with respect to the possible model and prior assumptions. 
In particular, as expected, the results are unaffected by the assumption of a different age prior (logarithmically in place of linearly uniform), while for the assumption of a linearly uniform metallicity prior (in place of a logarithmically uniform prior), the stellar metallicity estimates move to higher values, as expected, but the corresponding gradients are consistent with those of the reference case.
Different SFHs or formation redshift do not affect the resulting metallicity gradients at all.
Figure~\ref{fig:sens} shows that age gradients are poorly constrained in the reference case, and the results obtained in any of the test cases are fully consistent with them.  Results obtained considering templates that were derived from prolonged SFH models are not reported in this figure because the assumption of a prolonged initial star formation event corresponds to a smaller range of values of the age prior, and a direct comparison with the other cases is not possible. In Appendix B we show the joint age and metallicity probability distribution for the inner and outer zones of the sample that were obtained assuming a grid of templates derived from a top-hat SFH with a star formation timescale of $0.5$ Gyr. Their comparison with those of the reference case shows that the only change in the results is a shift towards older ages of both zones. The limiting maximum age corresponding to the redshift of the galaxies flattens the resulting age gradients. 
Therefore, systematic errors are subdominant compared to statistical errors.

To investigate the possible evolution of these galaxies, we now compare our results with those obtained at intermediate redshift and in the local Universe. In the local Universe, negative metallicity gradients are common within ETGs
\citep[e.g.][]{peletier1990ccd,idiart2003new,scott2009sauron,zibetti2020insights}. In particular, massive ETGs (M $ >10^{10} \ M_\odot$; e.g. \citealp{mehlert2003spatially,ogando2005observed,spolaor2009mass}) show negative metallicity gradients up to $-1.7$ dex/decade. Figure~\ref{fig:agemetgrad} compares the age and metallicity parameters in the inner zone and their gradients as found in our high-redshift sample with the typical values obtained for massive ETGs in cluster environments in the local Universe \citep{mehlert2003spatially}. \cite{mehlert2003spatially} measured the radial variation up to $1R_e$ of age and metallicity parameters for $35$ ETGs in the Coma cluster. The good match of the metallicity estimates (within $1 \sigma$ error) derived in two very distant cosmological epochs is remarkable and suggests that regardless of the evolution processes that occur in passive massive galaxies, the observed negative metallicity gradients must be kept from $z \sim 2$ towards $z=0$.
A purely passive evolution is consistent with the ages estimated in the inner zones and with the corresponding gradients in the two samples, from $z\sim 2$ towards $z = 0$, once the spread observed at $z = 0$ is accounted for. The
\cite{mehlert2003spatially} age, metallicity, and gradients are highly consistent with the determination of \cite{2006A&A...457..823S} for $37$ ETGs in the Coma cluster. This confirms the soundness of our comparison sample.

Our results are also consistent with those obtained for a sample of quiescent field galaxies at $z\sim0.8$ \citep{d2020inverse}, although a different environment was targeted. Because ground-based observations were used, D'Eugenio et a. (2020) were limited to analysing gradients on larger galaxy spatial scales than we probe.

To summarise, the negative metallicity gradients we found in the four high-redshift passive galaxies and their similarity with those retrieved 
in the local Universe suggest that the stellar populations of the passive galaxies are not redistributed over the last $10$ Gyr.
Our results support a scenario in which the main mechanism that determines the spatial distribution of the stellar population properties within passive galaxies is constrained in the first $3$ Gyr of the Universe. This evidence can easily be explained within the revised monolithic scenario described in Sect.~\ref{sec:intro}. 

\begin{figure*}
   \centering
   \subfloat[][\emph{}]
        {\includegraphics[width=0.45\textwidth]{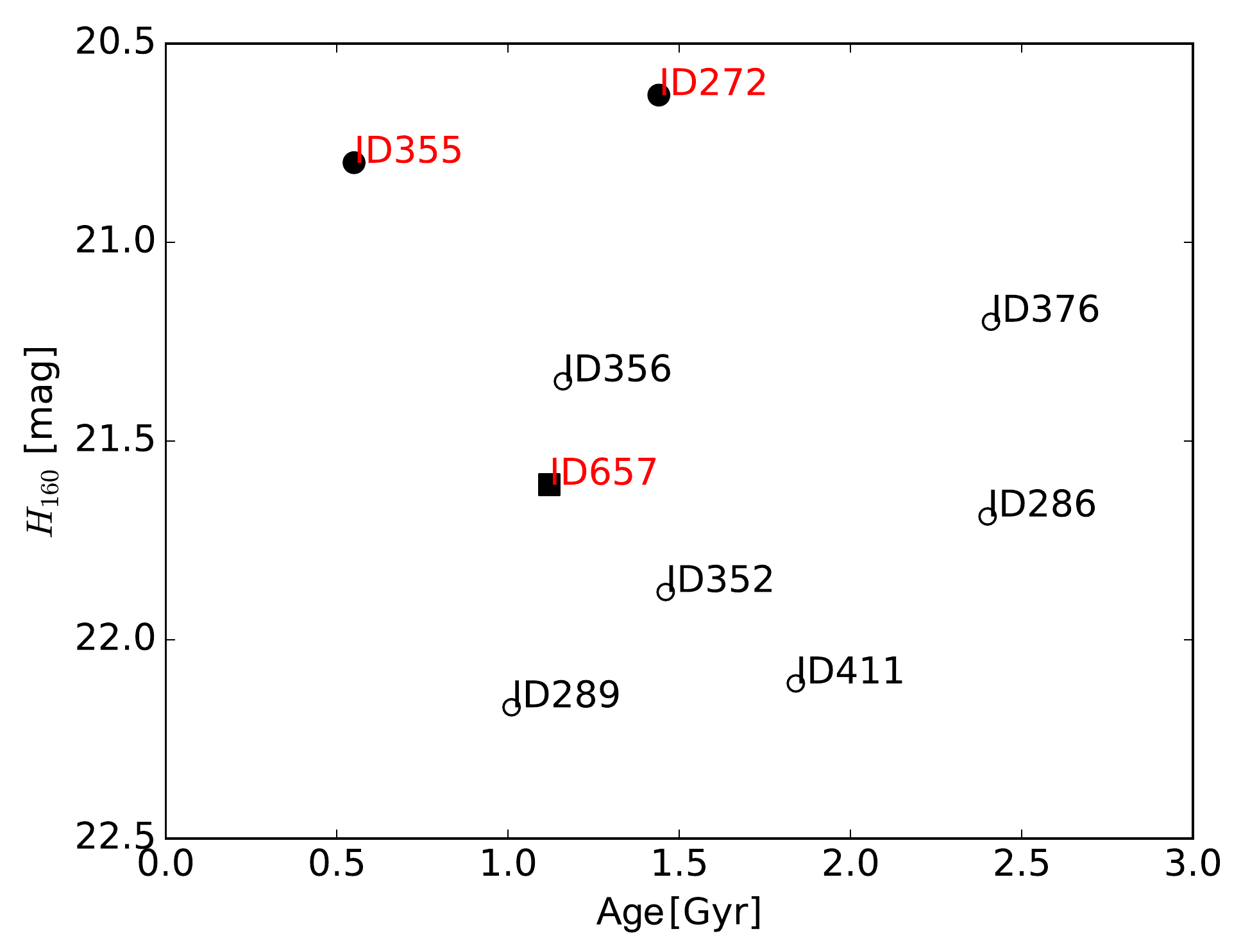}} \quad
        \subfloat[][\emph{}]
        {\includegraphics[width=0.45\textwidth]{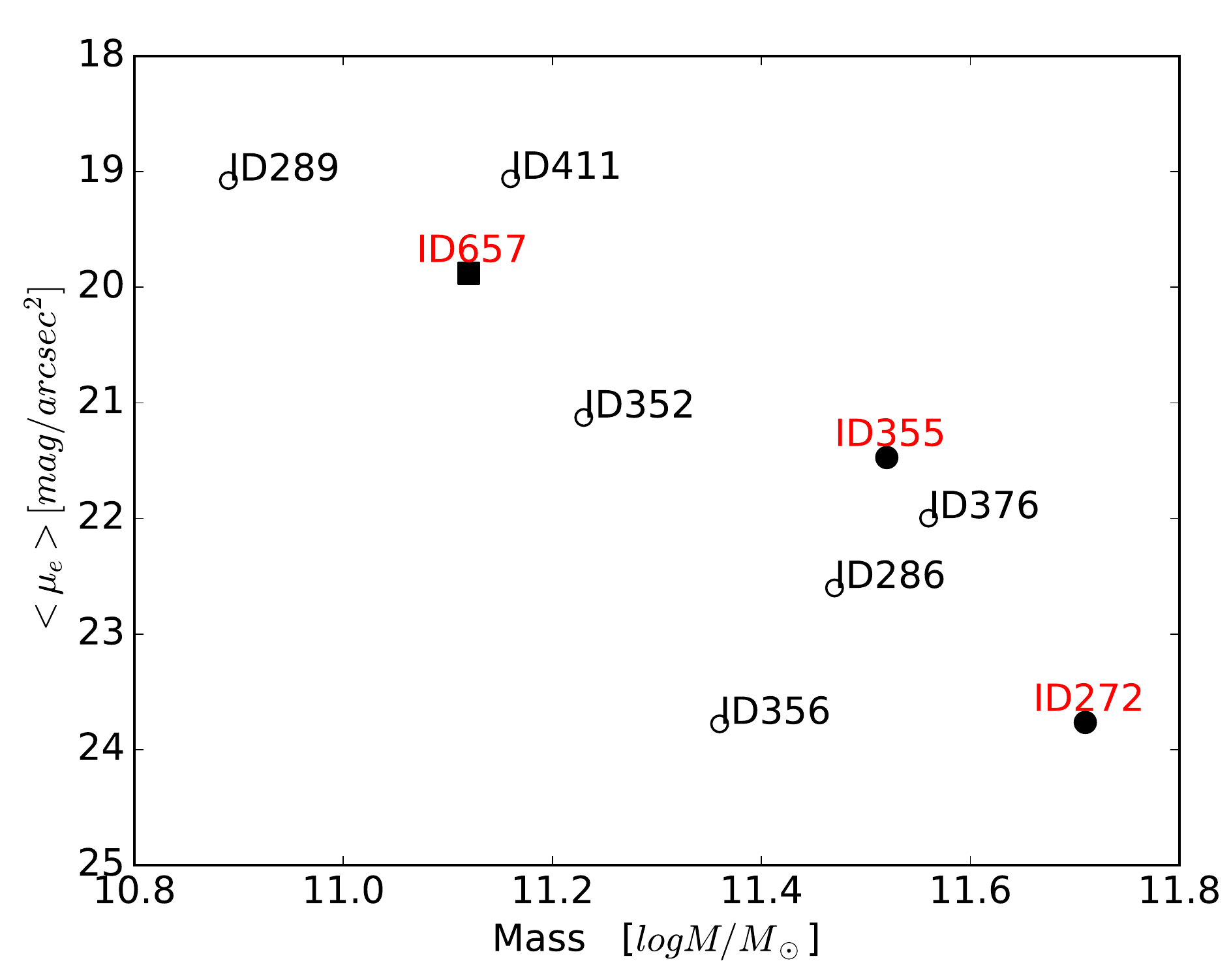}} \quad
        \subfloat[][\emph{}]
        {\includegraphics[width=0.45\textwidth]{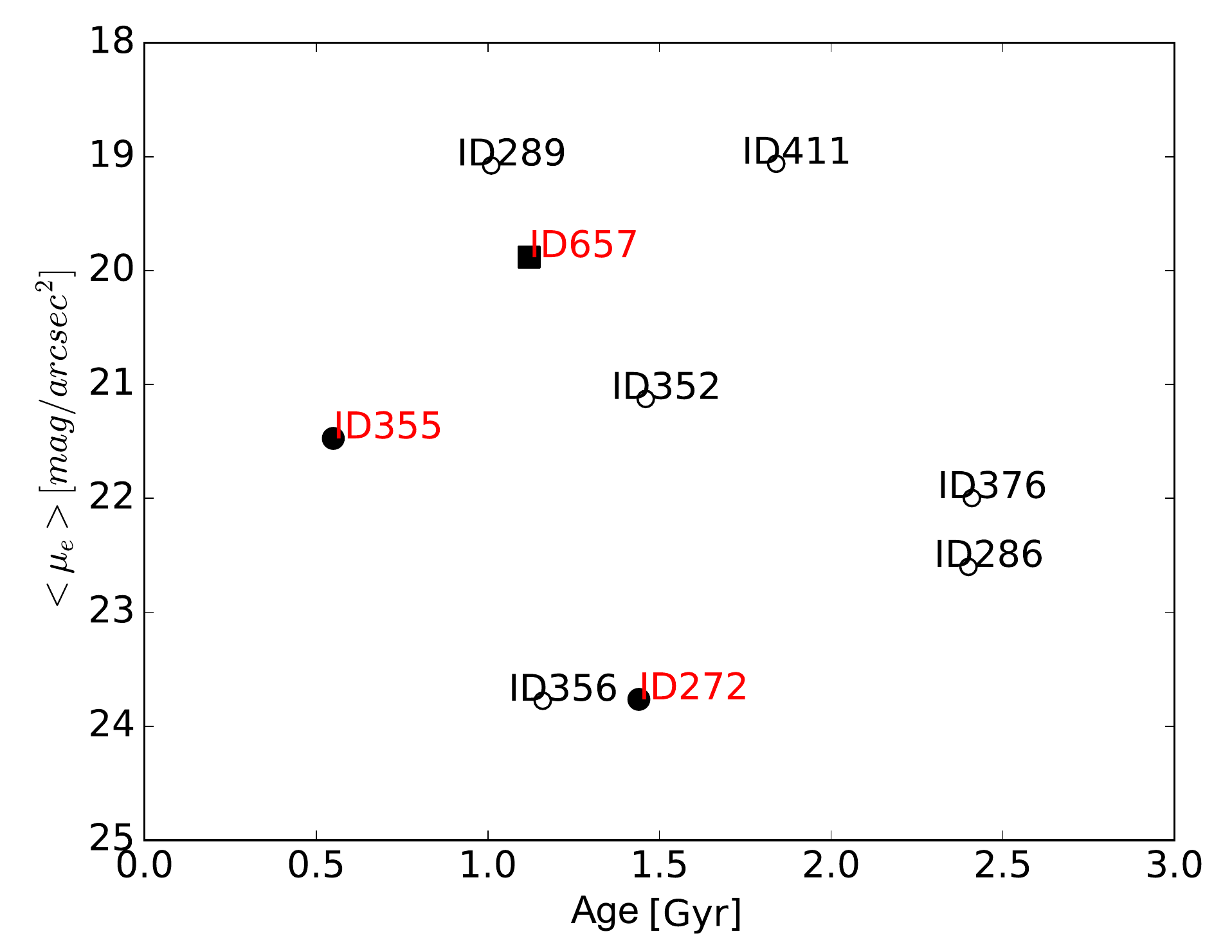}} 
   \caption{Scale relations for cluster galaxies. Panel $a$: $H_{160}$ magnitudes as a function of the ages of the quiescent galaxies of JKCS\,$041$. Panel $b$: Surface brightness within the effective radius as a function of the ages of the quiescent galaxies of JKCS\,$041$. Panel $c$: Surface brightness within the effective radius as a function of the mass of the quiescent galaxies of JKCS\,$041$. Data are taken from \cite{andreon2009jkcs}.}
              \label{fig:caveats}
    \end{figure*}

Our analysis, which is one of the first challenging attempts of measuring age and metallicity gradients in high-redshift galaxies, is surely limited by several factors: first of all, in the JKCS\,$041$ field of view,  we selected only the brightest and most massive three galaxies together with galaxy ID$657$ (Fig.~\ref{fig:magn}). Althoug ID657 is fainter, it has the right combination of surface brightness and dimensions in order to be successfully analysed. Nonetheless, the three cluster galaxies span a representative range of parameters such as age, surface brightness, or stellar mass, considering that they are all galaxies with M$>10^{11}$M$_\odot$ (see Fig.~\ref{fig:caveats}).
Moreover, the sample is small, therefore it is premature to extend our results to the entire class of passive galaxies at $z\sim 2$. Nevertheless, it is unlikely (although not impossible) that the global negative metallicity gradients confirmed in all four galaxies are due to chance, indicating that they are probably a common feature in high-redshift passive galaxies.

\section{Conclusions}
\label{sec:conclusions}

We presented the first estimates of the age and metallicity gradients based on spectroscopic data of a sample of $4$ high-redshift ($z \sim 2 $) spectroscopically passive galaxies in the JKCS\,$041$ field of view. The work is based on deep-grism slitless spectroscopic data taken with the G$141$ grism on HST/WFC$3$, which allowed us to spatially resolve starlight from galaxies with an apparent size of $\sim 1$ arcsecond.
The four galaxies for which the analysis was successful are part of a larger sample of $11$ galaxies, $3$ of which were immediately discarded due to their spectral contamination from other close objects within the field of view, while for the remaining four, the errors in age and metallicity were too large to derive any reliable conclusion.  Alhtough the number of successufully analysed galaxies is small, they span quite wide ranges of values in $R_e$ and $<\mu_e>,$ and they are only restricted in mass (M $> 10^{11}$M$_\odot$).

The four galaxies for which the analysis was successful (ID$355$, ID$272$, ID$657,$ and ID$64$) show negative metallicity gradients, although at only $1 \sigma$. 
Although the sample is small, the results we obtained and the comparison with metallicity and age gradients measured in the local Universe tend to support the revised monolithic scenario as the more probable paradigm for ETGs formation and evolution. We found metallicity gradients consistent with those confirmed at $z=0$ and $z=0.8$.

This work is the first study of stellar population properties gradients at high redshift based on spectroscopic data. It is therefore less limited by the age-metallicity degeneracy than the existing few works that are based on photometric data and colours \citep{guo2011color,gargiulo2012spatially}. Our analysis limits the effects of the possible presence of dust (selecting a narrow wavelength range for the analysis) and does not require the assumption of a particular age or metallicity
to measure the variation of the other. Moreover, our analysis has the advantage of being applicable to a larger sample of galaxies than analyses that exploit unlikely events, such as the gravitational lensing produced by a massive cluster on the line of sight \citep{jafariyazani2020resolved}.

Despite the validity of the analysis we carried out, this work is subject to some limitations that are mainly due to the very low spectral resolution of the HST slitless data. Moreover, because of the low S/N, the analysed data allow the measurement of gradients that exceed a minimum threshold value ($1.4$ dex/decade for metallicity, $1.2$ Gyr/decade for age), which limits the interpretation of the results in terms of galaxy formation scenarios. 
Despite the small sample size and the limitations affecting the analysis, the measurement of stellar population gradients in high-redshift galaxies using the HST grism has proven to be very promising in providing fundamental constraints on the galaxy formation scenarios. The analysis of larger samples, covering wider ranges of mass and sizes, is needed to derive a clear picture of the physical mechanisms that drive galaxy formation and evolution.

At the same time, our analysis has demonstrated the feasibility of an analysis like this even on the basis of very low spectroscopic resolution and at quite low S/N. This work has been successful not only in deriving the first reliable results, but also in tracing the way to identifying a set of spectroscopic data that is useful for this type of analysis.

The near future will offer many more opportunities through the advent of new facilities such as the James Webb Space Telescope, the Extremely Large Telescope and Euclid, which will provide not only the needed spatial resolution of about $0.1$ arcsec, but also intermediate spectral resolution and even the possibility of obtaining $2$D spectroscopic views of high-redshift galaxies.
Future instruments will thus enable the study of stellar population property distributions in large samples of galaxies that also include less massive objects and cover a wide range of angular dimensions. This will open a new era in the study of galaxy formation and evolution.

\begin{acknowledgements}
We thank the anonymous referee for carefully reading the manuscript and for the stimulating suggestions that helped to improve the paper.
\end{acknowledgements}

%
%
\bibliographystyle{aa}
\bibliography{aanda}

\begin{appendix} 
\section{Fit and joint probability distributions}

   \begin{figure*}
   \centering
   \includegraphics[width=0.95\textwidth]{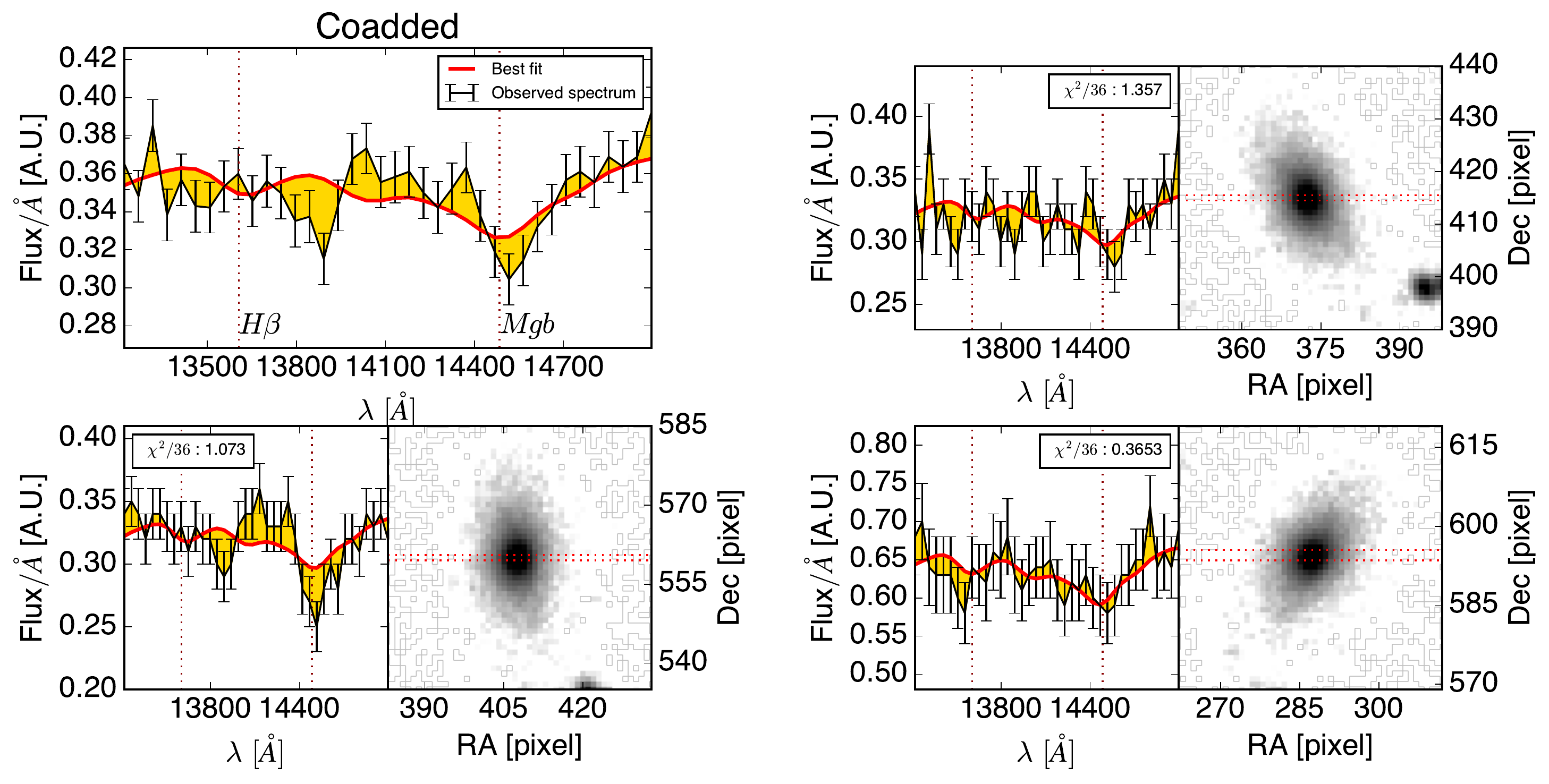}
      \caption{Extracted spectra and the corresponding best-fit templates in the inner zone of galaxy ID$272$. Upper left panel: Coadded spectrum of the inner zone of ID$272$ (line with errors) with the best-fit model in arbitrary units. Other panels: Single-visit observed spectrum of the inner zone of ID$272$ with the best-fit model and the corresponding F$160$W images with the related extraction.
              }
         \label{fig:central272}
   \end{figure*}
   

   \begin{figure*}
   \centering
   \includegraphics[width=0.95\textwidth]{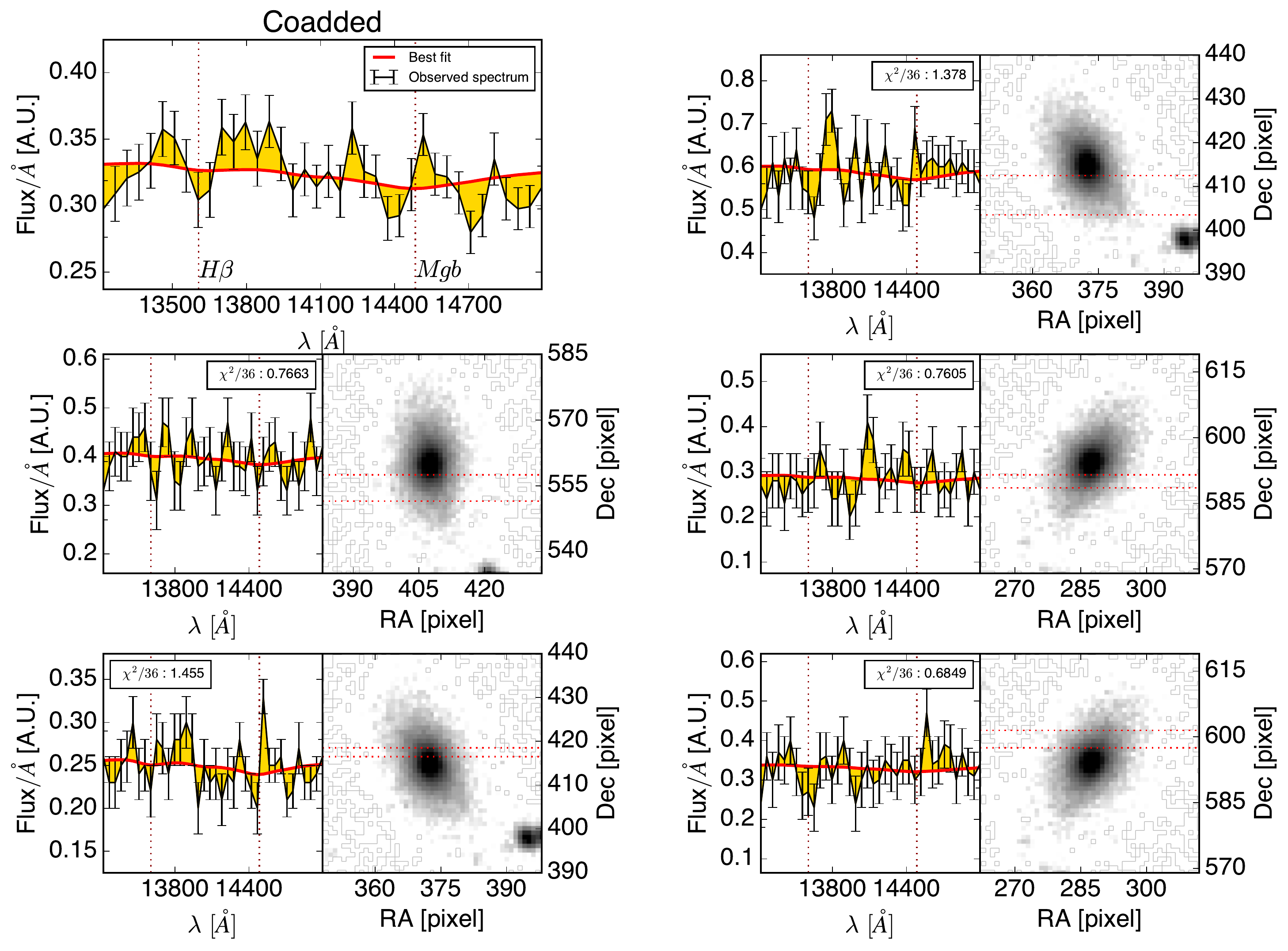}
      \caption{Same as Fig.~\ref{fig:central272} for the outer zone of ID$272$.
              }
         \label{fig:ext272}
   \end{figure*}

\begin{figure*}
   \centering
   \includegraphics[width=0.95\textwidth]{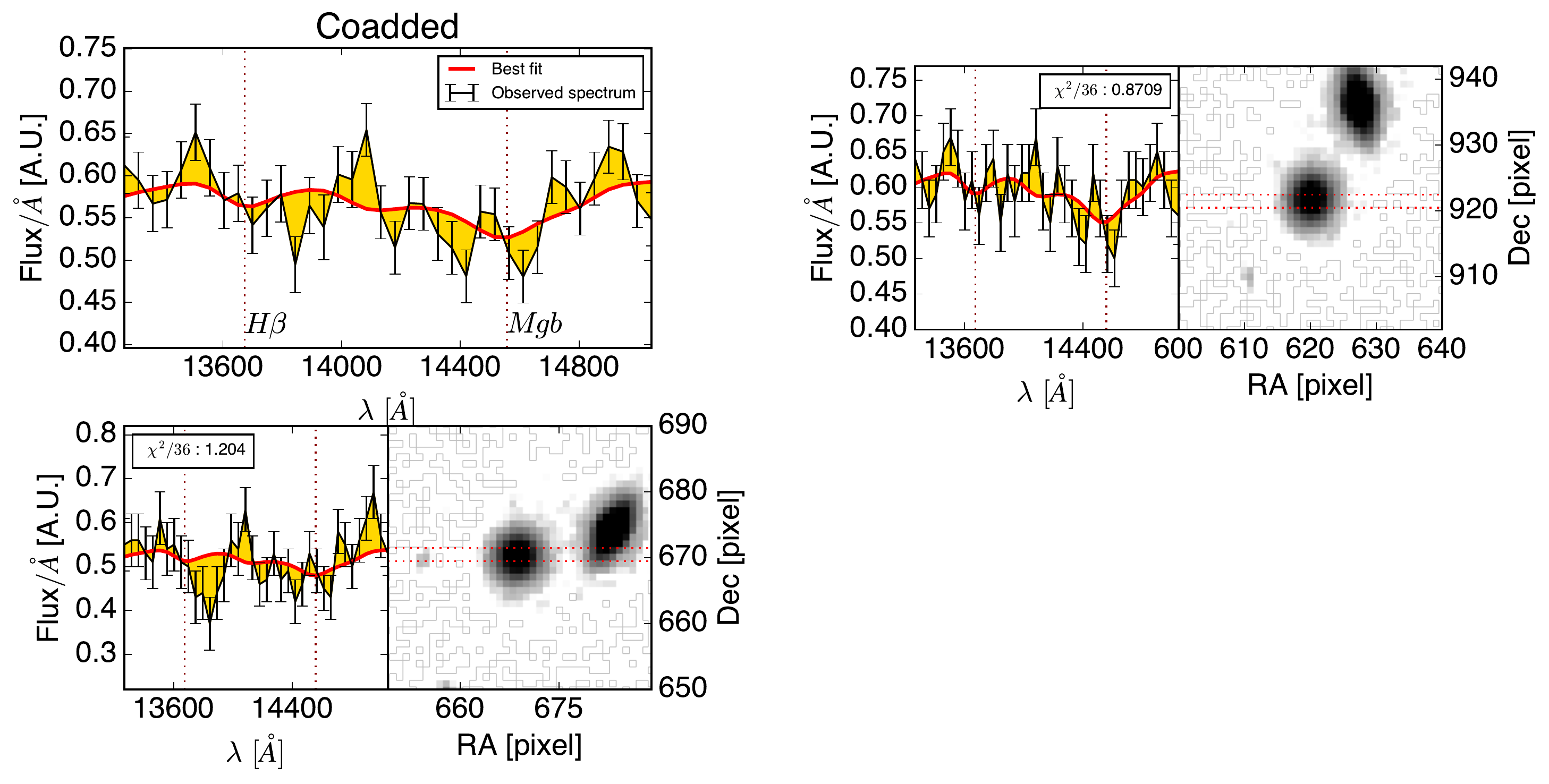}
      \caption{Same as Fig.~\ref{fig:central272} for the inner zone of ID$657$.
              }
         \label{fig:central657}
   \end{figure*}

\begin{figure*}
   \centering
   \includegraphics[width=0.95\textwidth]{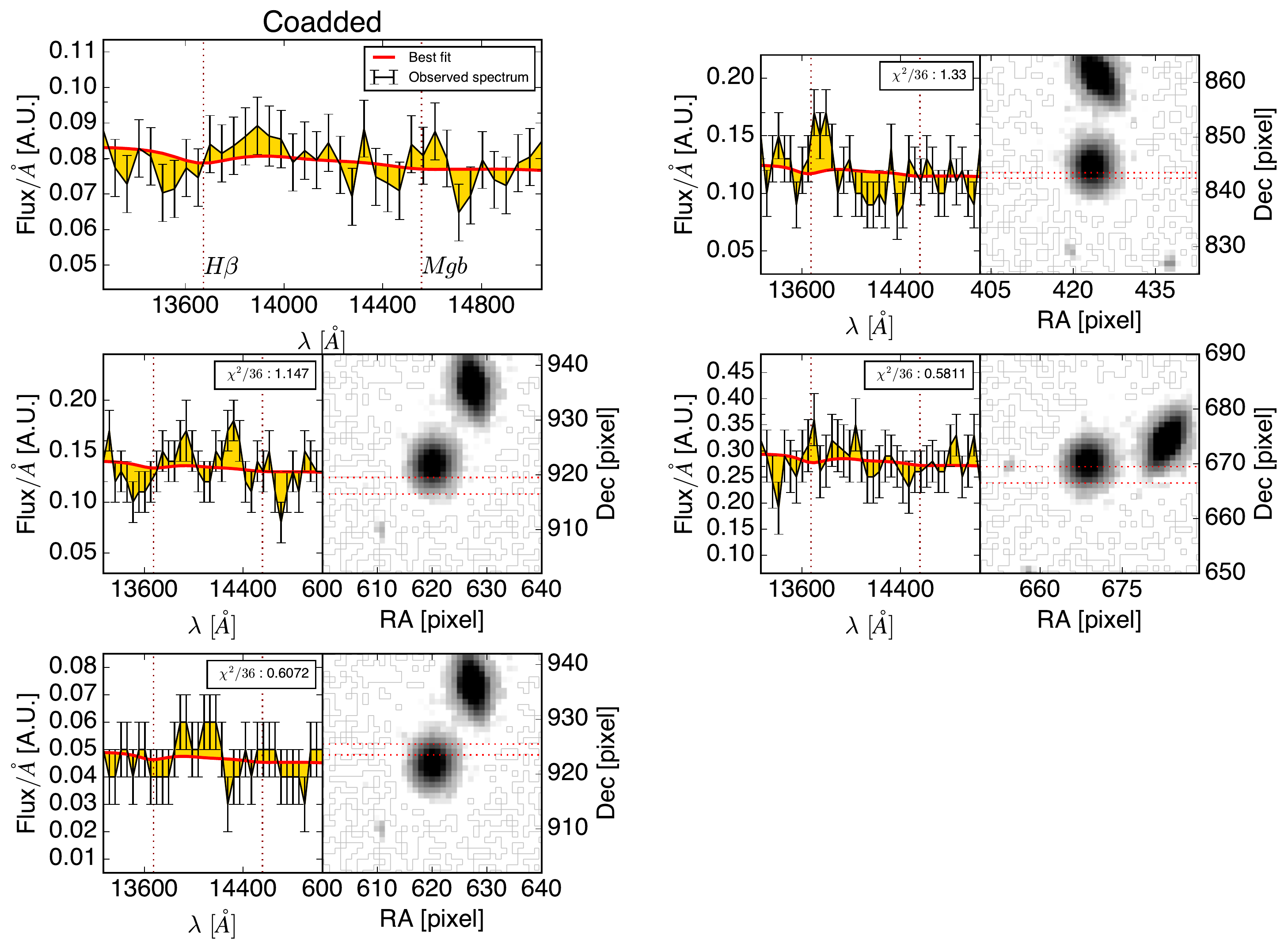}
      \caption{Same as Fig.~\ref{fig:central272} for the outer zone of ID$657$.
              }
         \label{fig:ext657}
   \end{figure*}

\begin{figure*}
   \centering
   \includegraphics[width=0.95\textwidth]{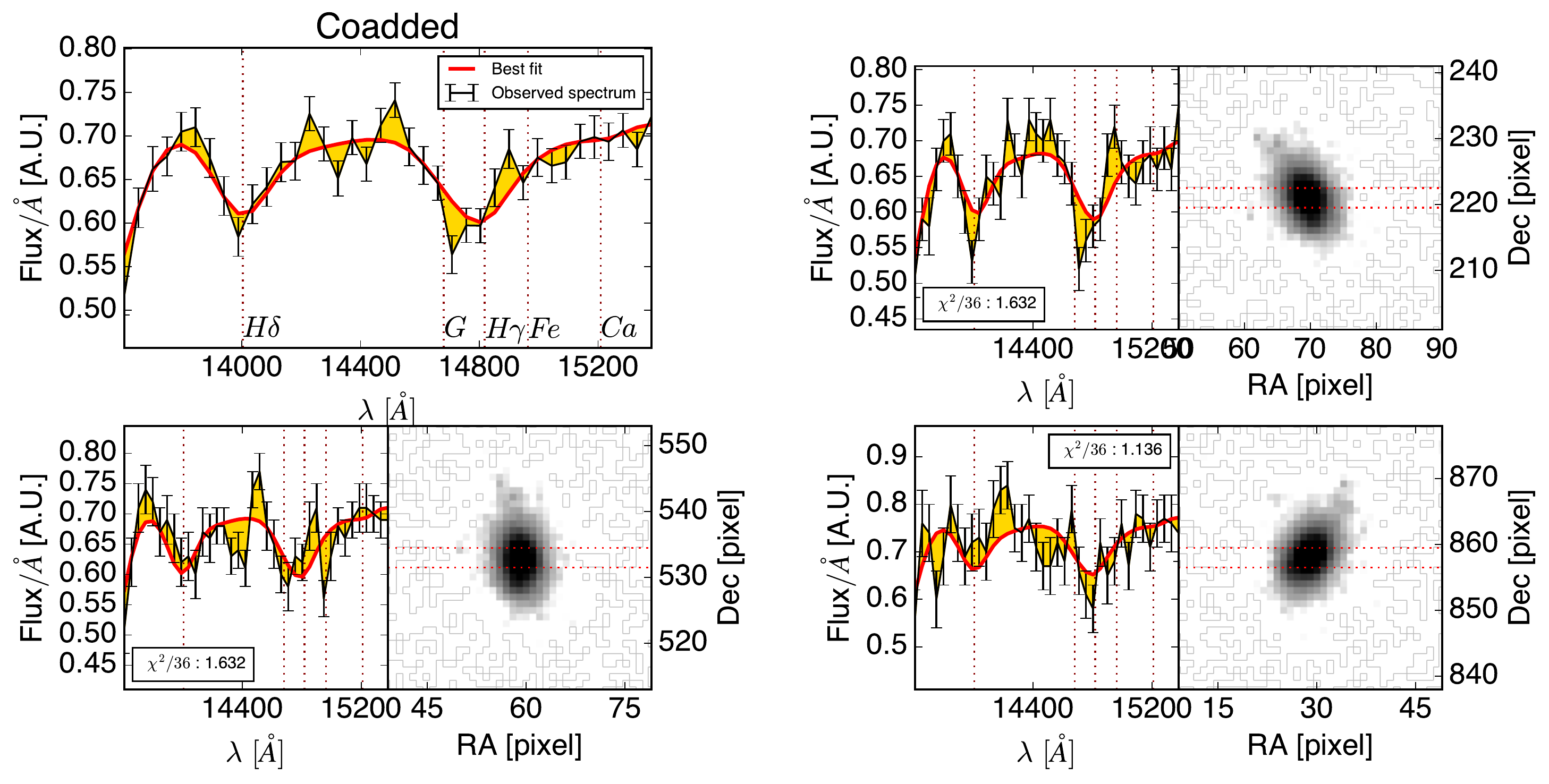}
      \caption{Same as Fig.~\ref{fig:central272} for the inner zone of ID$64$.
              }
         \label{fig:central64}
   \end{figure*}

\begin{figure*}
   \centering
   \includegraphics[width=0.95\textwidth]{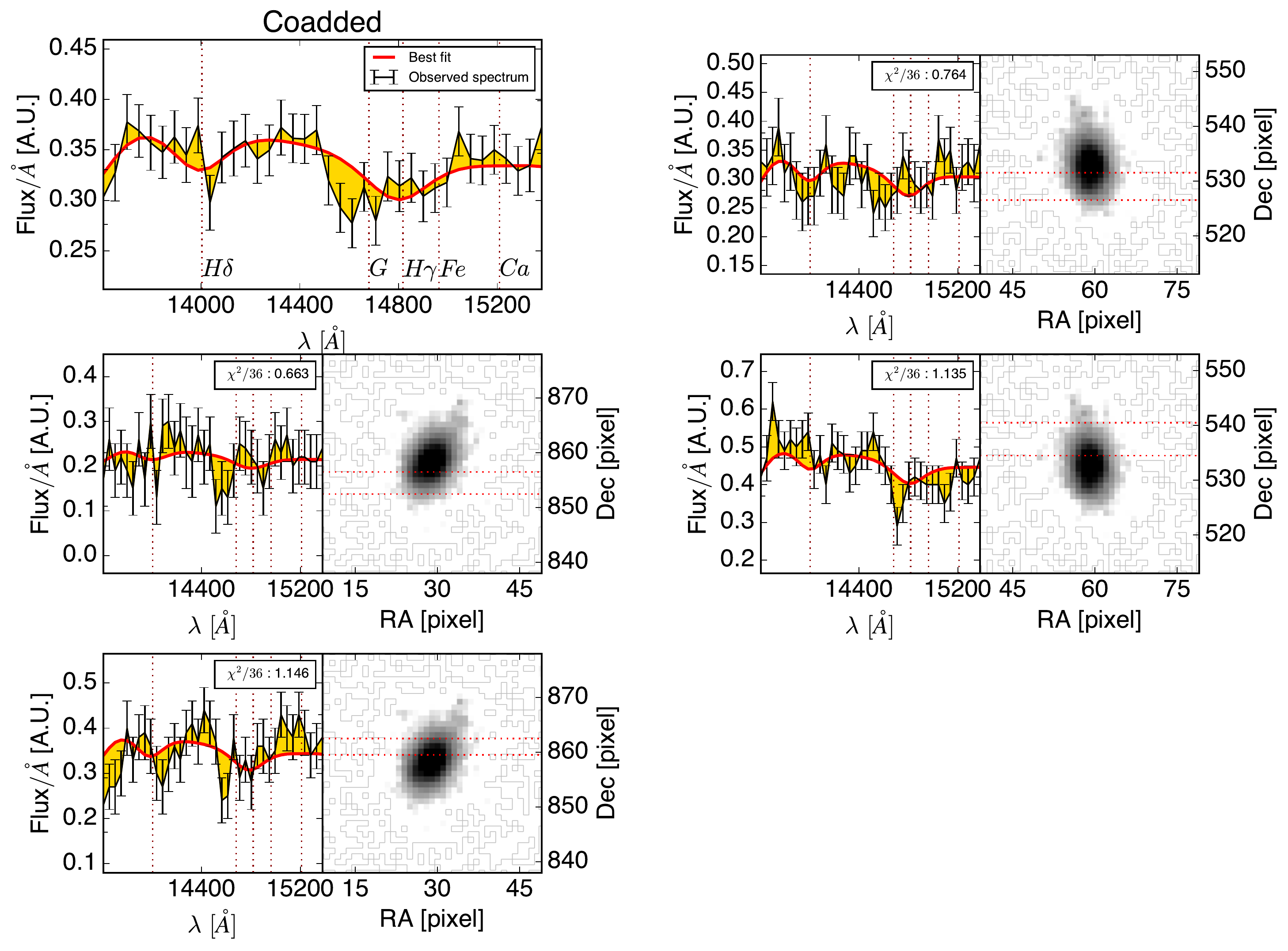}
      \caption{Same as Fig.~\ref{fig:central272} for the outer zone of ID$64$.
              }
         \label{fig:ext64}
   \end{figure*}

\begin{figure}
   \centering
   \includegraphics[width=0.5\textwidth]{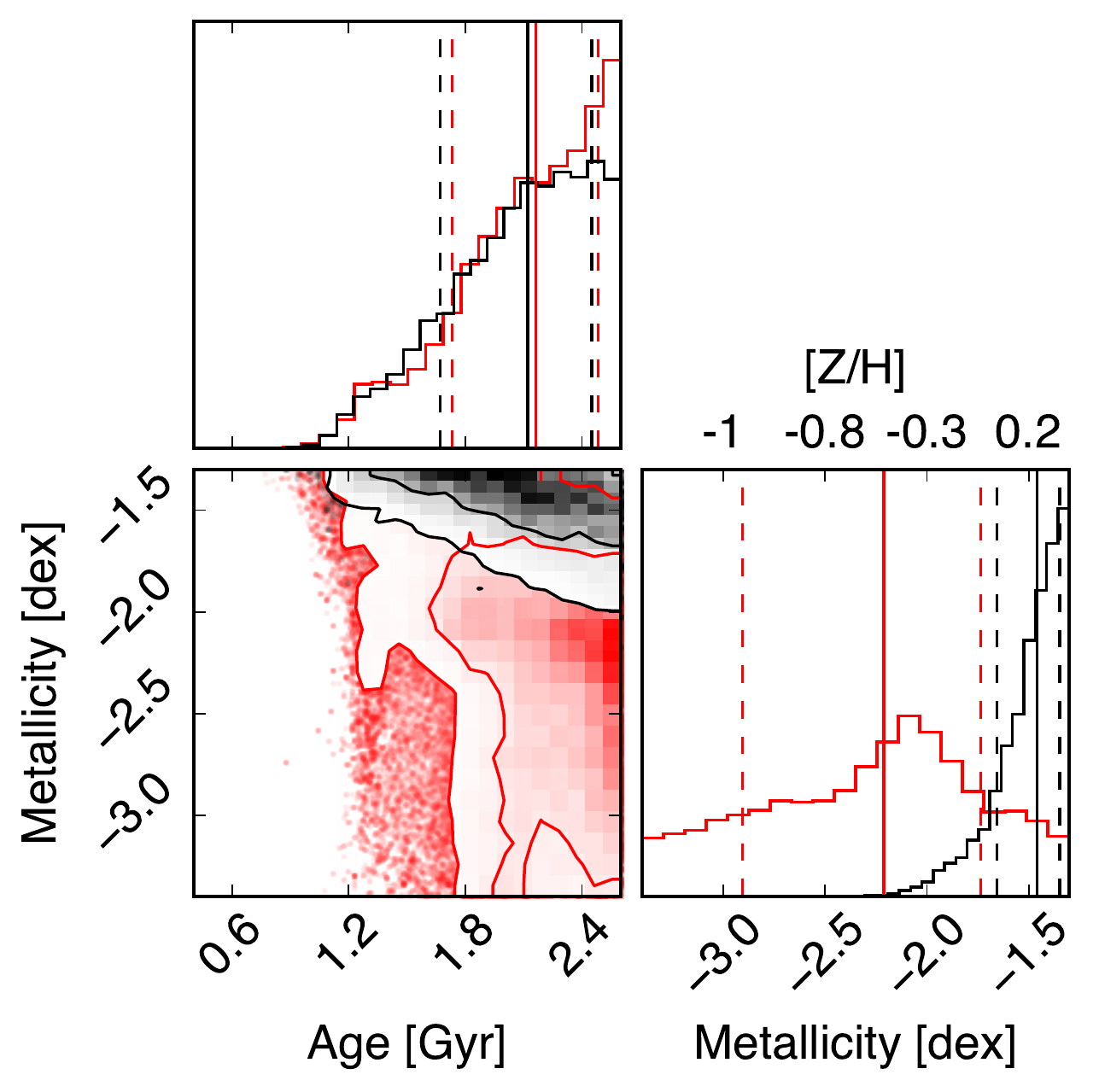}
      \caption{Combined age and metallicity distributions obtained for ID$272$. Lower left panel: Joint probability distribution of age and metallicity of the inner (black) and outer (red) zones of ID$272$. Contours are at 68\% and 95\% probability.
      Upper left panel: Marginalised probability for age. Lower right panel: Marginalised probability for metallicity. Median and 16\%\ and 84\% intervals are indicated by solid and dashed lines, respectively.
              }
         \label{fig:combined272}
   \end{figure}

\begin{figure}
   \centering
   \includegraphics[width=0.5\textwidth]{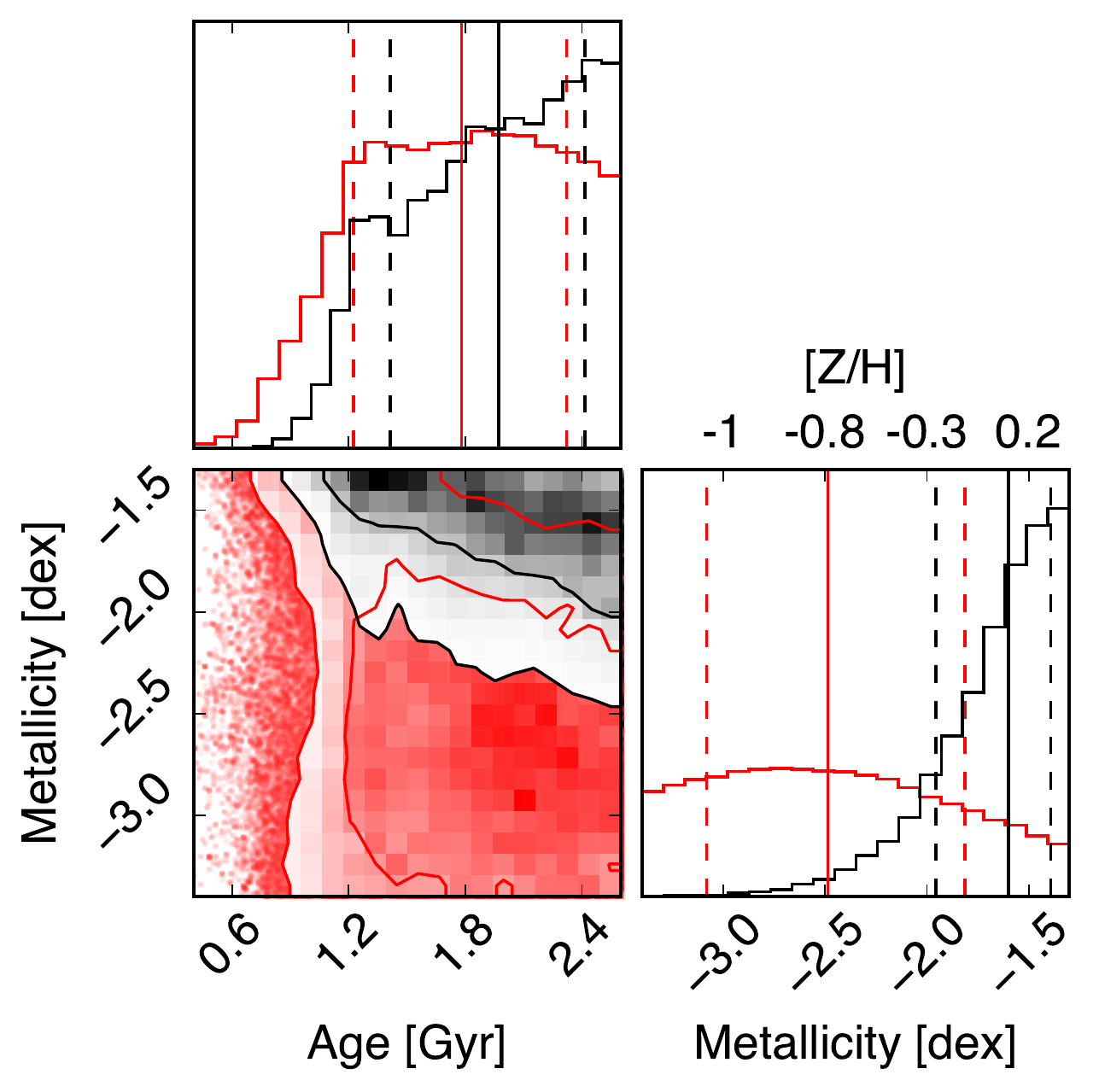}
      \caption{Same as Fig.~\ref{fig:combined272} for ID$657$.
              }
         \label{fig:combined657}
   \end{figure}

\begin{figure}
   \centering
   \includegraphics[width=0.5\textwidth]{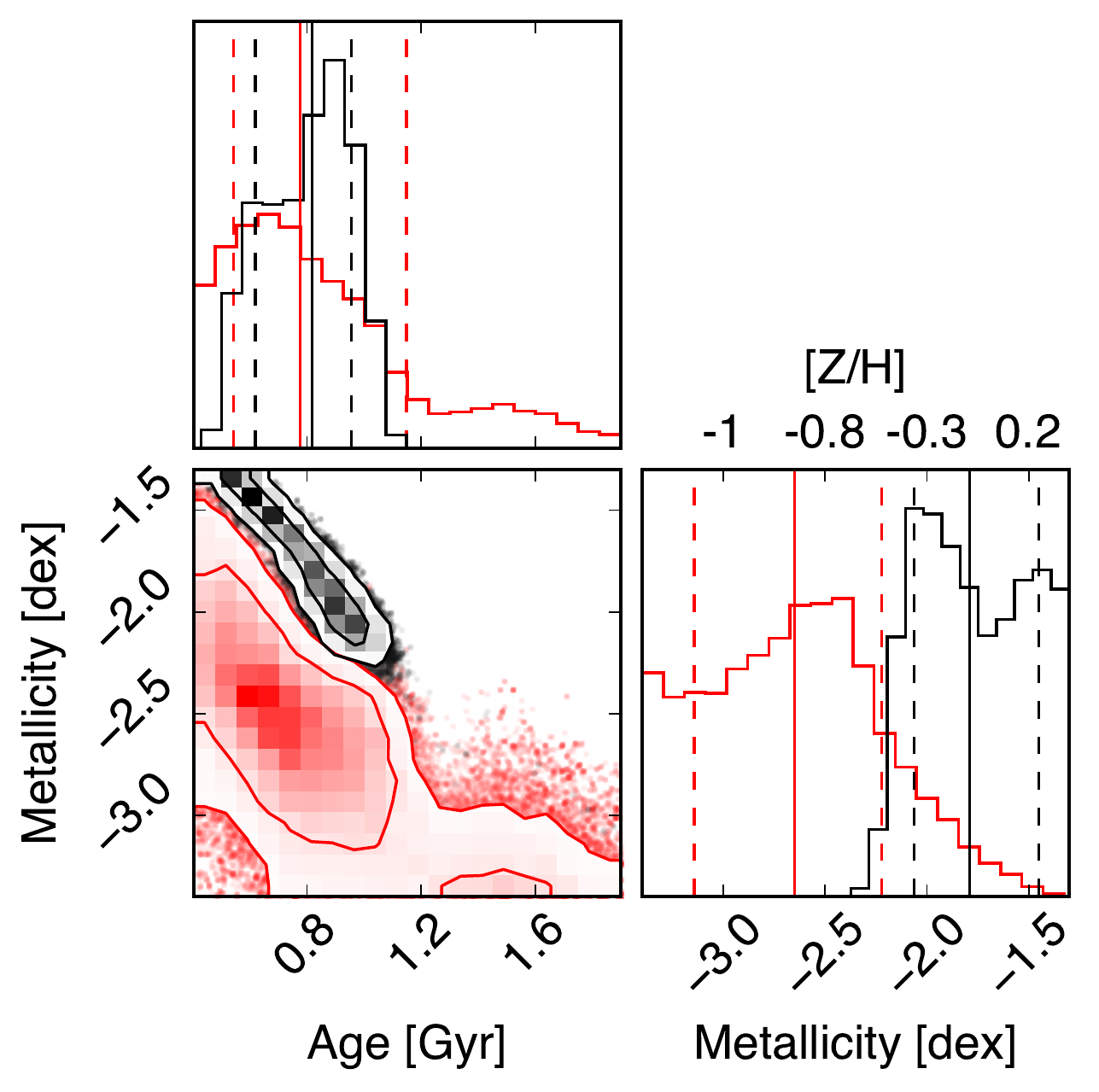}
      \caption{Same as Fig.~\ref{fig:combined272} for ID$64$.
              }
         \label{fig:combined64}
   \end{figure}

\begin{figure*}
   \centering
   \includegraphics[width = 0.8\textwidth]{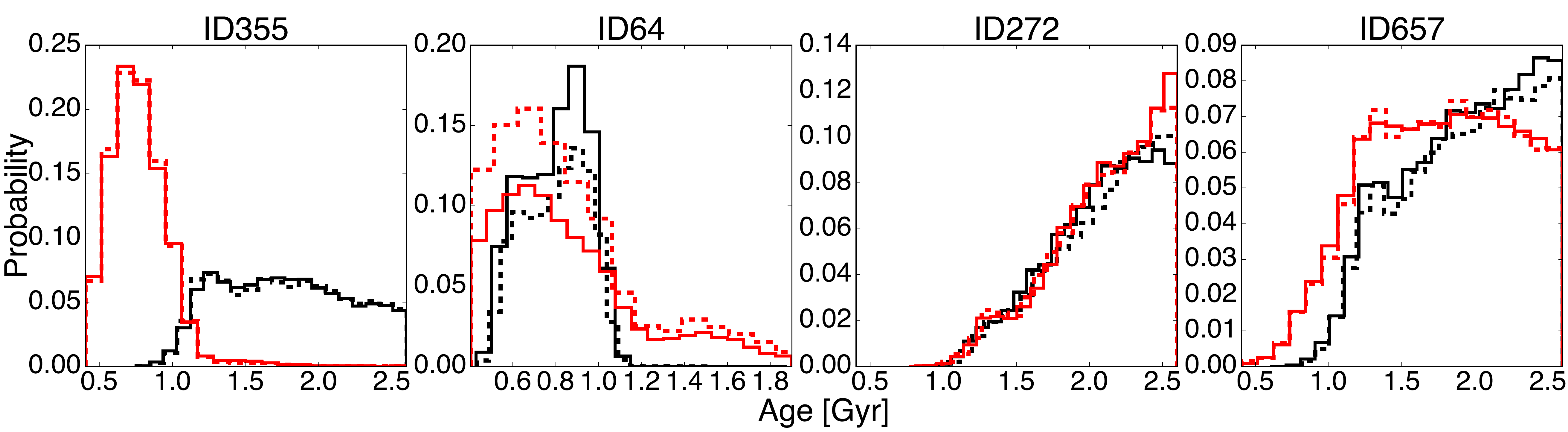}
   \caption{Comparison between the marginalised probability distribution of age in the case $A_V = 0$ (solid lines) and $0 < A_V < 0.7$ (dashed lines) of the dust attenuation in the inner (black) and outer (red) zones for the $\text{four}$ galaxies. }
              \label{fig:mainvsdustage}
    \end{figure*}

\begin{figure*}
   \centering
   \includegraphics[width = 0.8\textwidth]{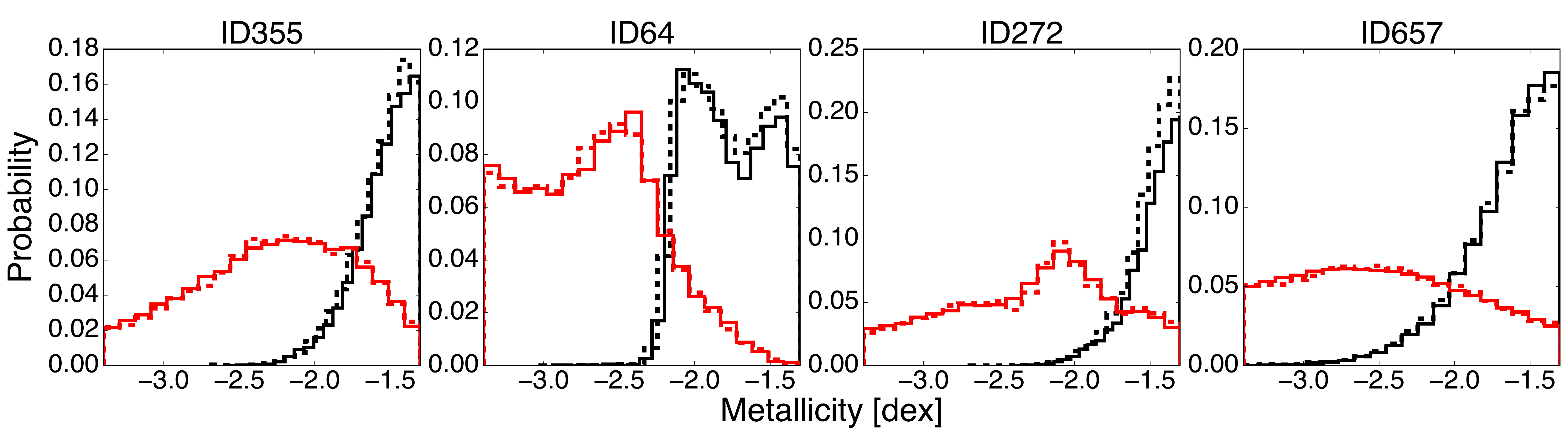}
   \caption{Same as Fig.~\ref{fig:mainvsdustage} for metallicity. }
              \label{fig:mainvsdustmetal}
    \end{figure*}

In this appendix we show the fit and joint probability distributions of the extracted spectra in the inner and outer zones for each galaxy. Figures from~\ref{fig:central272} to~\ref{fig:ext64} show the fit results for the inner and outer zone of ID$272$, ID$657,$ and ID$64$. Figures from~\ref{fig:combined272} to~\ref{fig:combined64} show the joint probability distributions for ID$272$, ID$657,$ and ID$64$.\\
We verified whether the estimates of age and metallicity depend on dust attenuation $A_{V}$ assuming the \cite{calzetti2000dust} law as a third parameter in the analysis. Considering the passive nature of our sample, we adopted values between $0 <A_{V}< 0.7$.  Figures~\ref{fig:mainvsdustage} and~\ref{fig:mainvsdustmetal} show the comparison between marginalised probability distributions of age and metallicity, respectively, of the extracted spectra in the inner and outer zones for each galaxy in the case $A(V) = 0$ and $0 < A(V) < 0.7$. These figures demonstrate that dust attenuation does not change the age and metallicity probability distribution for these galaxies appreciably.

    \FloatBarrier

\section{Joint probability distributions with extended SFH}


Even though the sampled galaxies were selected on the basis of their spectroscopic quiescency (i.e. $\tau_{best} < 0.1$ Gyr for exponentially declining SFHs) and their high redshift implies a very short star formation timescale, we performed the analysis assuming a top-hat SFH with a star formation timescale of $0.5$ Gyr and $1.5$ Gyr in order to access the dependence of the results on the SFH model assumptions.
Figure~\ref{fig:mainvssfh} shows the joint probability distributions for ID$355$, ID$272$, ID$64,$ and ID$657$ in the reference case and with a top-hat SFH with a star formation timescale of $0.5$ Gyr. The ages reported in the prolonged SFH case are mass-weighted mean ages. This figure demonstrates that a prolonged SFH does not change the metallicity probability distributions, while the age distributions are shifted to older ages, as expected.
Regarding the more prolonged SFH (i.e. with a timescale of $1.5$ Gyr), the corresponding templates do not fit the observed spectra at all (i.e. the $\chi^2_{red}$ of the best-fitting model is larger than $3$). 

\begin{minipage}{\textwidth}
    \centering
    \vspace{1cm}
    \includegraphics[width = 0.9\textwidth]{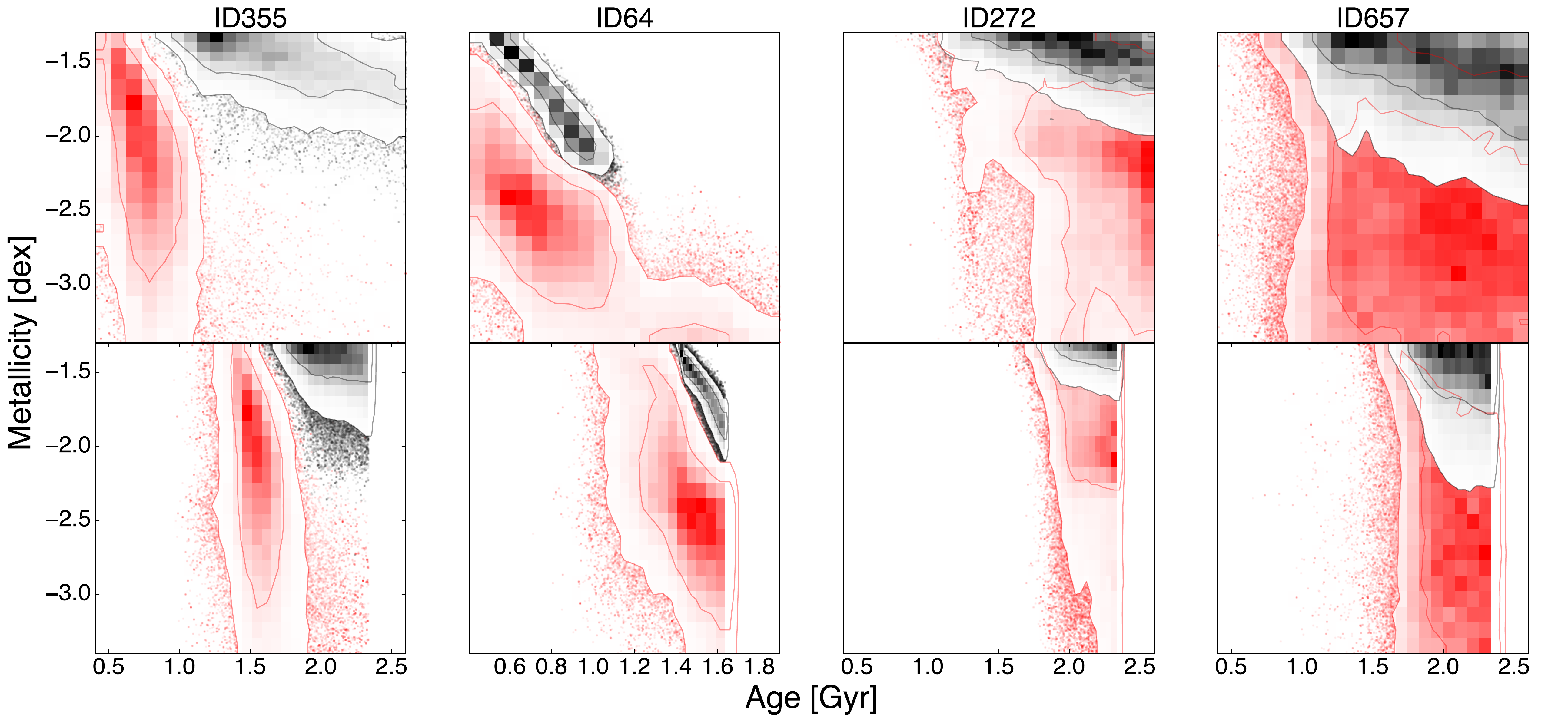}
    \captionof{figure}{Comparison between the joint probability distribution of age and metallicity of the inner (black) and outer (red) zones in the reference case (upper panels) and with a top-hat SFH with a star formation timescale of $0.5$ Gyr (lower panels) for the $\text{four}$ galaxies. Contours are at $68\%$ and $95\%$ probability.}
    \label{fig:mainvssfh}
\end{minipage}

\end{appendix}
\end{document}